\documentclass[12pt]{article}
\usepackage{amsmath,amssymb}
\usepackage{amsthm}
\usepackage{enumerate}
\usepackage[dvipdfmx]{graphicx}
\usepackage{cite}
\usepackage{mathrsfs}
\numberwithin{equation}{section}
\usepackage[dvipdfmx]{hyperref}

\newcommand{\qtq}[1]{\quad\text{#1}\quad}
\newcommand{\er}[1]{Eq.~\eqref{#1}}
\newcommand{\ers}[1]{Eqs.~\eqref{#1}}
\newcommand{\hc}{\text{h.c.}}
\newcommand{\hph}{\hphantom}
\renewcommand{\b}{\bar}
\renewcommand{\d}{\dot}
\newcommand{\pd}[2]{\frac{\partial{#1}}{\partial{#2}}}
\newcommand{\sr}{\sqrt}
\newcommand{\df}{\dfrac}
\newcommand{\der}{\partial}
\renewcommand{\(}{\left(}
\renewcommand{\)}{\right)}
\newcommand{\dg}{\dagger}
\newcommand{\bmx}{\left(\begin{matrix}}
\newcommand{\emx}{\end{matrix}\right)}
\newcommand{\mtx}[1]{\bmx #1 \emx}
\usepackage[top=2.828cm,bottom=2.828cm, left=3cm,right=3cm]{geometry}
\begin{document}
\begin{titlepage}
\hfill 
\vspace{-1em}
\def\thefootnote{\fnsymbol{footnote}}%
   \def\@makefnmark{\hbox
\       to\z@{$\m@th^{\@thefnmark}$\hss}}%
 \vspace{3em}
 \begin{center}%
  {\Large 
Higher derivative three-form gauge theories
\\ 
and their supersymmetric extension
  \par
   }%
 \vspace{1.5em}
  {\large
Muneto Nitta\footnote{nitta@phys-h.keio.ac.jp}
and 
Ryo Yokokura\footnote{ryokokur@keio.jp}
   \par
}
 \vspace{1.5em} 
{\small \it
Department of Physics \& 
Research and Education Center for Natural Sciences,
Keio University, Hiyoshi 4-1-1, Yokohama, Kanagawa 223-8521, Japan
 \par
}
\vspace{1.5em} 
   {\large
    }
 \end{center}
 \par
\vspace{1.5em}
\begin{abstract}
We investigate three-form gauge theories with higher derivative interactions and their supersymmetric extensions in four space-time dimensions. 
For the bosonic three-form gauge theories, 
we show that derivatives on the field strength of the 3-form gauge field yield a tachyon as far as the Lagrangian contains a quadratic kinetic term,
while such the term with opposite sign gives rise to a ghost. 
We confirm that there is neither a tachyon nor 
a ghost when all higher derivative terms 
are given by functions of the field strength.
For this ghost/tachyon-free Lagrangian, we determine the 
boundary term necessary for the consistency between the equation of motion and energy-momentum tensor.
For supersymmetric extensions, 
we present ghost/tachyon-free higher derivative interactions of arbitrary order of the field strength and corresponding boundary terms as well. 
\end{abstract}
\end{titlepage}
 \setcounter{footnote}{0}%
\def\thefootnote{$*$\arabic{footnote}}%
   \def\@makefnmark{\hbox
       to\z@{$\m@th^{\@thefnmark}$\hss}}% 
\tableofcontents
\newpage
\section{Introduction}

3-form gauge theories in four-dimensional (4D) spacetime have been studied extensively in the past decades.
They were first considered in the context of quantum chromodynamics
(QCD) to describe a long-range confinement force 
between quarks~\cite{Aurilia:1977jz,Aurilia:1978dw}.
Furthermore, a 3-form gauge field can be regarded to provide  
an effective description of a Chern--Simons 3-form  in 
Yang--Mills theories~\cite{Luscher:1978rn,Aurilia:1978dw}, 
in particular in the context of the $U(1)$ problem~\cite{Aurilia:1980jz,Hata:1980hn}
and the strong CP problem~\cite{Dvali:2005an,Dvali:2004tma,Dvali:2005zk}.
In cosmology, a 3-form gauge field was used for dynamical neutralization of the cosmological constant~\cite{Aurilia:1980xj,Hawking:1984hk,Brown:1987dd,Brown:1988kg,Duncan:1989ug,Duff:1989ah,Bousso:2000xa,Wu:2007ht}, quintessence~\cite{Kaloper:2008qs}, 
inflationary models~\cite{Kaloper:2008fb,Kaloper:2011jz,Marchesano:2014mla,Kaloper:2014zba,Kaloper:2016fbr,DAmico:2017cda}.
A relation between the 3-form gauge theories 
and condensed matter physics
was also discussed, see e.g.~\cite{Ansoldi:1995by}.
A supersymmetric (SUSY) extension of the 3-form gauge fields was first formulated in Ref.~\cite{Gates:1980ay}. 
The SUSY 3-form gauge fields naturally appear in superstring theory and M-theory, therefore they were studied extensively with various applications: supergravity (SUGRA)~\cite{Gates:1980az,Buchbinder:1988tj,Binetruy:1996xw}
(see Refs.~\cite{Gates:1983nr,Buchbinder:1998qv} as a review), 
St\"uckelberg coupling~\cite{Groh:2012tf,Hartong:2009az,Becker:2016xgv,Aoki:2016rfz},
topological coupling~\cite{Dudas:2014pva,Becker:2016xgv,Yokokura:2016xcf},
coupling with a membrane~\cite{Ovrut:1997ur,Kuzenko:2017vil,Bandos:2018gjp},
alternative formulation of old-minimal 
SUGRA~\cite{Ovrut:1997ur,Kuzenko:2005wh,Farakos:2016hly,Farakos:2017jme},
gaugino condensation in SUSY Yang--Mills 
theories~\cite{Binetruy:1995hq,Binetruy:1995ta},
the cosmological constant problem~\cite{Farakos:2016hly},
SUSY breaking~\cite{Farakos:2016hly,Buchbinder:2017vnb},
string effective theories~\cite{Bielleman:2015ina,Valenzuela:2016yny,Montero:2017yja,Farakos:2017jme}, 
and inflationary models~\cite{Dudas:2014pva,Yamada:2018nsk}.
Complex 3-form gauge theories were also considered 
in Refs.~\cite{Farakos:2017jme,Kuzenko:2017vil,Bandos:2018gjp}.

One of the characterizations of $p$-form gauge fields is their couplings to extended objects.
As 1-form and 2-form gauge fields 
can be electrically 
coupled to a particle and string, 
respectively in 4D spacetime, 
a 3-form gauge field can be electrically 
coupled to a membrane~\cite{Aurilia:1977jz}.
Since membranes and 3-form gauge fields naturally arise 
in string theory and M-theory as fundamental degrees of freedom,
3-form gauge fields in 4D spacetime appear as 4D compactification of these theories. 
Another characteristic feature of the 3-form gauge field is its coupling to scalar fields.
Since a field strength of the 3-form gauge field is 
a 4-form, the 3-form gauge field can be topologically 
coupled to a pseudo-scalar 
field~\cite{Aurilia:1980jz,Curtright:1980yj,Dvali:2005an,Kaloper:2008fb}.
It was pointed out that 
the topological coupling generates a potential (e.g., mass term)
for the pseudo-scalar field while preserving a shift symmetry
of the pseudo-scalar field
in an action~\cite{Kaloper:2008qs,Kaloper:2008fb}.
This mechanism was applied to inflationary 
models~\cite{Marchesano:2014mla,Franco:2014hsa,Kaloper:2016fbr,DAmico:2017cda}, quintessence~\cite{Kaloper:2008qs}, 
the strong CP problem~\cite{Dvali:2005an,Dvali:2004tma,Dvali:2005zk} and so on.

Apart from many applications, the 3-form gauge field would have some theoretical and fundamental subtleties which we have to discuss carefully, 
compared with other $p$-form gauge theories. 
The 3-from gauge field with canonical (quadratic derivative) kinetic term
has no dynamical degrees of freedom, and hence 
it is classically dual to a constant term,
analogous to an electromagnetic field in 2D spacetime.
Nevertheless, there are several merits to consider 
a 3-form gauge field itself rather than merely a constant.
In fact, it was shown that a 3-form gauge field is 
inequivalent to a constant at a quantum level~\cite{Duff:1980qv}.

In contrast to other $p$-form gauge theories, Lagrangians of the 3-form gauge theories generally should include boundary terms~\cite{Brown:1987dd, Brown:1988kg}. 
If the boundary term were missing, the functional variation of the 3-form gauge field at the boundary would not vanish, and consequently 
the energy-momentum tensor of the 3-form gauge field 
would be inconsistent with its 
equation of motion
(EOM)~\cite{Duncan:1989ug,Duff:1989ah}.
This situation is the same with 
the $\theta$-term in 2D electromagnetism, 
where the $\theta$-term is needed if one considers a 
non-trivial field strength in the bulk.

As mentioned above, 3-form gauge theories are often 
considered to describe infrared (low-energy or long-range) effective theories,
such as a Chern--Simons 3-form in QCD, and compactifications
 of string theory and M-theory.
Since effective theories inevitably include nonrenormalizable interactions which depend on an ultraviolet cutoff parameter,
it is natural to consider higher derivative corrections 
to the 3-form gauge fields.
Since the Lagrangian should be gauge invariant, 
the nonrenormalizable interactions may be described in terms of the field strength of the 3-form gauge field
rather than the 3-form gauge field itself.
In particular, higher derivative corrections in 3-form gauge 
theories provide a pseudo-scalar field with a non-trivial 
potential~\cite{Dvali:2005an,Dvali:2005zk,DAmico:2017cda}
such as a cosine-type potential~\cite{Dvali:2005an}.
This is in contrast to the case of the quadratic derivative term 
(kinetic term) giving rise to only a mass term for the pseudo-scalar field.

\begin{table}[t]
{\small
\begin{tabular}[t]{|c|c|c|}
\hline
 & Bosonic & SUSY
\\
\hline
Instability 
& 
$-$ 
&
$-$
(Cf.~chiral superfield~\cite{Antoniadis:2007xc})
\\
\hline
2nd (bulk)
&
\cite{Aurilia:1977jz}
&
\cite{Gates:1980ay}
\\
\hline
2nd (boundary)
&
\cite{Brown:1987dd}
&
\cite{Groh:2012tf,Farakos:2017jme}
\\
\hline
H.D.~(bulk)
&
 Ghost/tachyon-free~\cite{Dvali:2005an,Dvali:2005zk}
&
Ghost/tachyon-free (4th order)~\cite{Dudas:2014pva}
\\
&
&
Ghost/tachyon (all order, chiral)~\cite{Ciupke:2016agp}
\\
&
&
Ghost/tachyon-free (all order)~$-$
\\
\hline
H.D.~(boundary)
&
$-$
&
$-$
\\
\hline
\end{tabular} 
}\caption{{\small Higher derivative Lagrangians 
of 3-form gauge theories.
``2nd'' and ``H.D.'' imply second order canonical term
and higher derivative term, respectively.
``$-$'' implies that it has not been done and 
will be done in this paper.
The reason why we list chiral superfield is that 3-form gauge theories can be formulated in terms of a chiral superfield.
}
}
\label{tab}
\end{table}

Construction of consistent higher derivative 3-form 
gauge theories consists of three procedures 
for both bosonic and SUSY cases: 
\begin{enumerate}[(1)]
 \item  identifying unstable modes such as ghosts and tachyons, 
\item constructing bulk Lagrangians free from unstable modes,  
\item  determining boundary Lagrangian corresponding to the bulk.
\end{enumerate}
Which have already been done and which have not been are summarized 
in Table \ref{tab}.

\begin{enumerate}[(1)]
 \item 
In general, higher derivative interactions give rise to additional degrees of freedom, which cause instabilities.
For higher derivative theories of a scalar field $\phi$, 
such an instability is known as the Ostrogradsky's ghost~\cite{Ostrogradsky:1850fid, Woodard:2006nt}.
A sufficient condition for the absence of such a ghost is that 
the Lagrangians should be written by a function of 
the first order derivative (such as $\der \phi$), 
but not by the higher order derivatives 
than the first order (such as $\der \der\phi$).
However, in the case of the 3-form gauge theories, 
the existence or absence of such unwanted degrees of freedom 
is not known in general. 
\item 
The examples of the higher derivative Lagrangians 
without unstable modes is known in bosonic 3-form gauge 
theories. One of sufficient conditions free from unstable modes 
is that the Lagrangian 
consists of arbitrary function of the field strength 
of a 3-form gauge field~\cite{Dvali:2005an,Dvali:2005zk,DAmico:2017cda}.
However, this is not a necessary condition; 
Even when there exist derivative terms of the 
field strength of a 3-form gauge field~\cite{Klinkhamer:2016jrt,
Klinkhamer:2016zzh,Klinkhamer:2016lgk}, 
there are no unstable modes, 
if the canonical kinetic term 
is absent or has the wrong sign~\cite{Klinkhamer:2017nfb}.
On the other hand, in contrast to the bosonic case, 
the only known example of a higher derivative term free from 
unstable modes in the SUSY 3-form gauge theories 
is a four derivative term~\cite{Dudas:2014pva}.

One of the most characteristic features of the SUSY 3-form gauge fields,  in contrast to non-SUSY cases, is that there exist dynamical degrees of freedom of bosons and fermions even at on-shell, 
which are superpartners of the 3-form gauge field.
In higher derivative theories,
one should be careful with 
unstable modes 
originated from 
higher derivative terms of the dynamical degrees of freedom.

When we construct higher derivative extensions of SUSY
3-form gauge theories, we can use  
higher derivative Lagrangians for a chiral superfield 
because the field strength of the 3-form gauge field can be 
expressed in terms of a chiral superfield.
This is because a 3-form gauge field can be embedded into a vector component of a real superfield, and its 
field strength can be embedded into an auxiliary field component of a chiral superfield
 which is defined by 
the real superfield~\cite{Gates:1980ay}.
These are analogous to a vector superfield and a chiral superfield 
(gaugino superfield) for a SUSY electromagnetism, respectively.
A quartic order term of the field strength was 
described in Ref.~\cite{Dudas:2014pva} as mentioned above.
Although higher order terms of an auxiliary field were considered 
in Ref.~\cite{Ciupke:2016agp},
they contain the Ostrogradsky's ghost instability in general.
The most general ghost-free higher derivative terms for SUSY 3-form fields are still missing.

This is in contrast to the cases of chiral 
matter and vector superfields.
For the chiral superfields, 
ghost-free higher derivative Lagrangians were systematically 
given in Refs.~\cite{Khoury:2010gb,Khoury:2011da,Koehn:2012ar,Koehn:2012te}, which were later generalized in Ref.~\cite{Nitta:2014pwa}.
Before their constructions, 
it has been known that higher derivative terms in SUSY theories often encounter the so-called auxiliary field problem 
\cite{Gates:1995fx, Gates:1996cq, Gates:2000rp, Gates:2000gu, Gates:2000dq}: the action contains the terms with spacetime derivatives on auxiliary fields. 
In this case, the auxiliary terms cannot be eliminated by their EOM, 
which is the case for a Wess--Zumino term~\cite{Nemeschansky:1984cd,Nitta:2001rh} and Skyrme-like models~\cite{Bergshoeff:1984wb,Freyhult:2003zb}.
This auxiliary field problem usually comes up together 
with the higher derivative ghosts~\cite{Antoniadis:2007xc},
although such the ghost can be removed by introducing 
a non-dynamical gauge field 
(the ghostbuster mechanism) \cite{Fujimori:2016udq,Fujimori:2017rcc}.
The higher derivative terms free from these problems
given in Refs.~\cite{Khoury:2010gb,Khoury:2011da,Koehn:2012ar,Koehn:2012te,Nitta:2014pwa}
were applied to many topics such as 
low-energy effective theories~\cite{Buchbinder:1994iw,Buchbinder:1994xq,Banin:2006db,Kuzenko:2014ypa},  
SUGRA~\cite{Koehn:2012ar,Farakos:2012qu}, 
SUSY extension~\cite{Khoury:2011da,Queiruga:2016yzd} 
of Galileons~\cite{Nicolis:2008in},  
ghost condensation~\cite{Khoury:2010gb,Koehn:2012te}, 
the Dirac--Born--Infeld (DBI) inflation~\cite{Sasaki:2012ka}, 
flattening of the inflaton potential~\cite{Aoki:2014pna,Aoki:2015eba},
baby Skyrme models~\cite{Adam:2013awa,Adam:2011hj,Nitta:2014pwa,Nitta:2015uba,Bolognesi:2014ova,Queiruga:2016jqu,Queiruga:2018nph},
Skyrme-like models~\cite{Gudnason:2015ryh,Gudnason:2016iex,Queiruga:2015xka}, 
 solitons~\cite{Nitta:2014pwa,Nitta:2015uba,Queiruga:2017blc,Eto:2012qda}, 
nonlinear realizations~\cite{Nitta:2014fca}, 
SUSY breaking in modulated vacua~\cite{Nitta:2017yuf,Nitta:2017mgk}, 
and a formulation of a liberated SUGRA~\cite{Farakos:2018sgq}. 
For the vector superfield case, 
ghost-free higher derivative actions were considered in the context of a correction to a scalar potential in SUGRA~\cite{Cecotti:1986jy},
SUSY Euler--Heisenberg model~\cite{Farakos:2012qu,Antoniadis:2007xc,Dudas:2015vka}, nonlinearly self-dual actions~\cite{Kuzenko:2000uh,Kuzenko:2002vk}, 
and the DBI action~\cite{Cecotti:1986gb,Bagger:1996wp}.
The most general ghost-free action of an arbitrary order of the field strength was achieved in Ref.~\cite{Fujimori:2017kyi}.
Later, higher derivative theories were applied to
the Fayet--Iliopoulos term without gauged R-symmetry~\cite{Cribiori:2017laj,Aldabergenov:2018nzd,Kuzenko:2018jlz}, and inflationary models~\cite{Aldabergenov:2017hvp,Abe:2018plc}.
\item 
The boundary terms are also needed corresponding to 
higher derivative terms 
for the consistency between EOM and the energy-momentum tensor.
However, in both the bosonic and SUSY cases, 
they have not been explicitly presented. 
\end{enumerate}

In this paper, we give all constructions of 
3-form gauge theories missing in Table~\ref{tab}.
First, in the bosonic case, we show that higher derivative terms given by 
derivatives on the field strength may cause a tachyon
as far as the 
canonical kinetic term exists.
If such higher derivative terms are absent, 
there are no additional degrees of freedom.
We further argue that there are no additional degrees of 
freedom if the higher derivative terms are given by 
functions of the field strength but not of derivatives 
of the field strength.
All the previously known examples in Refs.~\cite{Dvali:2005an,Dvali:2005zk,Kaloper:2011jz,Kaloper:2016fbr,DAmico:2017cda} 
fall into this class.

Second, we give the most general higher derivative Lagrangian 
including an arbitrary order of the field strength 
without tachyons as well as ghosts in the SUSY case.
Since the field strength of the 3-form gauge field 
can be embedded into the auxiliary field of a chiral superfield, 
we can construct a ghost-free higher derivative system of 3-form gauge fields by a Lagrangian with an arbitrary function of the auxiliary field.
This Lagrangian is obtained by choosing 
ghost-free sector of the Lagrangian in Ref.~\cite{Ciupke:2016agp}.
We show that this Lagrangian is ghost-free and tachyon-free in the bosonic sector. 

Third, we determine the boundary terms for the higher derivative Lagrangians given by arbitrary functions of the field strength in both
the bosonic and SUSY cases.
These boundary terms are determined by requiring the 
condition that a functional variation of the Lagrangian
vanishes at the boundary.

This paper is organized as follows.
In Sec.~\ref{bch}, 
we first review a 3-form gauge theory with 
a quadratic kinetic term and the role of the boundary term.
We then argue that a Lagrangian 
with derivatives on the 
field strength gives rise to 
a tachyon by using an example.
We then discuss 
a tachyon-free higher derivative Lagrangian,
and its the boundary term.
We further confirm that our boundary term 
gives us an energy-momentum tensor consistent with EOM.
In Sec.~\ref{sch}, we discuss SUSY extension of 
ghost/tachyon-free higher derivative 3-form gauge theories.
First, we review 3-form gauge theories with 
4D ${\cal N}=1$ 
global SUSY with a quadratic kinetic term
and the corresponding boundary term.
Second, 
we propose the ghost/tachyon-free 
Lagrangian for SUSY 3-form gauge theories.
We then determine the boundary term for the ghost/tachyon-free 
Lagrangian by requiring
that a superspace functional variation
 at a boundary should vanish.
Section~\ref{sum} is devoted to a summary and discussion.
In Appendix~\ref{notation}, we summarize our notation.
In Appendix~\ref{bd}, we review a dual formulation of
a 3-form gauge theory.
In Appendix~\ref{sd}, we also review 
a dual formulation of 
a 3-form gauge theory in SUSY field theories.
In Appendix~\ref{sgf}, 
we discuss an auxiliary field method for 
the ghost/tachyon-free Lagrangian proposed in Sec.~\ref{sh}. 

We use the notation of the textbook~\cite{Wess:1992cp}.

%%%%%%%%%%%%%%%%%%%%%%%%%%%%%%%%%%%%%%%%%%%%%%%
\section{3-form gauge theories}
\label{bch}

In this section, we consider 3-form gauge theories in 4D.
First, we review a 3-form gauge theory with a canonical kinetic term with 
a corresponding boundary term.
Second, we argue that a higher derivative term 
given by a derivative of the field strength of the 3-form gauge field 
gives rise to a tachyon in the presence of 
the canonical kinetic term.
We also show that higher derivative Lagrangians are  
tachyon-free if the higher derivative terms are given by 
a function of the field strength but not of a function of the 
derivatives of the field strength.
Finally, we consider the necessity of the 
boundary term for the higher derivative term, and 
we specify a higher derivative extension of the boundary term.

\subsection{3-form gauge theory with canonical kinetic term}
\label{bc}

Here, we review a 3-form gauge field, its field strength, and 
gauge invariant Lagrangian with a canonical kinetic term
with a boundary term.
This review part is mainly based on 
Refs.~\cite{Brown:1987dd,Brown:1988kg,Duncan:1989ug}.

%%%%%%%%%
\subsubsection{3-form gauge field}
A 3-form gauge field is a third-rank antisymmetric tensor field
which is transformed by a 2-form antisymmetric tensor local
parameter $\xi_{mn}$ as follows:
\begin{equation}
\delta_3 C_{mnp}
 = 
\der_m \xi_{np}
+
\der_n \xi_{pm}
+
\der_p \xi_{mn}.
\label{5}
\end{equation}
Here, $\delta_3$ denotes an infinitesimal gauge transformation
 of the 3-form gauge field.
The field strength of the 3-form gauge field is introduced 
as follows:
\begin{equation}
 F_{mnpq}
= 
\der_m C_{npq}
-
\der_n C_{mpq}
+
\der_p C_{mnq}
-
\der_q C_{mnp}.
\end{equation}
The field strength is invariant under the gauge transformation of 
the 3-form:
\begin{equation}
\delta_3 F_{mnpq} = 0. 
\end{equation}
Note that the field strength can be written by 
using totally anti-symmetric tensor $\epsilon_{mnpq}$:
\begin{equation}
 F_{mnpq} 
= 
-\df{1}{4!}\epsilon_{mnpq}\epsilon^{rstu}F_{rstu}
\label{180725.1731}
\end{equation}
because $F_{mnpq}$ is totally antisymmetric tensor in 4D.
It is convenient to define the Hodge dual of the 
field strength $F$ as follows:
\begin{equation}
 F := \df{1}{4!} \epsilon^{mnpq} F_{mnpq}.
\label{180724.1657}
\end{equation}
We can write $F_{mnpq}$ in terms of $F$ as
\begin{equation}
 F_{mnpq} = -\epsilon_{mnpq} F.
\label{180725.1822}
\end{equation}

\subsubsection{Lagrangian with canonical kinetic term: bulk part}
In the following, we give a Lagrangian with a quadratic derivative
term.
As we will explain below, the Lagrangian gives us a constant term.
As a merit, we can describe the cosmological 
constant in terms of a gauge symmetry~\cite{Duncan:1989ug,Brown:1987dd,Brown:1988kg,Bousso:2000xa}.
A quadratic derivative Lagrangian of the 3-form gauge field is 
given by
\begin{equation}
 {\cal L}_\text{kin.} 
=
-\df{1}{2\cdot 4!} F^{mnpq} F_{mnpq}
+ \df{1}{3!}\der^m (C^{npq}F_{mnpq}).
\label{3}
\end{equation}
The first term is a quadratic derivative term,
which we call the canonical kinetic term.
The second term is a boundary term corresponding to the kinetic term, which is necessary for the consistency as described below.
We impose the gauge invariant boundary condition for the
3-form gauge field:
\begin{equation}
  F|_{\text{bound.}} = -c,
\label{180725.1524}
\end{equation}
where $c$ is a real constant, and the minus sign is just a convention.
The symbol $|_{\text{bound.}}$ denotes the value at the boundary.
We further impose that the functional variation 
of the field strength at the boundary is zero: 
\begin{equation}
 \delta F|_\text{bound.}=0,
\label{180727.2110}
\end{equation}
which we will use to discuss the boundary term.
Note that the Lagrangian in \er{3} can be rewritten by using $F$ 
as
\begin{equation}
{\cal L}_\text{kin.} = 
+\df{1}{2} F^2
- \df{1}{3!}\der^m (C^{npq}\epsilon_{mnpq} F).
\end{equation}
The Lagrangian written in term of $F$ will be useful when 
we consider higher derivative extensions.

The variation of the Lagrangian in \er{3} by the 3-form gauge field 
gives us the EOM of the 3-form gauge field:
\begin{equation}
 \der^m F_{mnpq} =0, \qtq{or equivalently} 
\der^m F= 0.
\end{equation}
This can be solved as
\begin{equation}
 F_{mnpq} = \epsilon_{mnpq} c,
\end{equation}
where the constant $c$ is determined by the 
boundary condition in \er{180725.1524}.
We will use them below.

\subsubsection{Lagrangian with canonical kinetic term: boundary part}
Now, we review a role of the boundary term.
The boundary term is necessary in order that the variation 
of the kinetic term at the boundary vanishes:
the variation of the Lagrangian by the 3-form gauge field is 
\begin{equation}
\begin{split}
 \delta {\cal L}_\text{kin.}
&
=
-\df{1}{3!} (\der^m \delta C^{npq}) F_{mnpq}
+ \df{1}{3!}\der^m (\delta C^{npq}F_{mnpq})
\\
&
=
-\df{1}{3!} \der^m (\delta C^{npq} F_{mnpq})
+ \df{1}{3!}\der^m (\delta C^{npq}F_{mnpq})
\\
&\quad
+ 
\df{1}{3!}\delta C^{npq} \der^m  F_{mnpq}.
\end{split}
\label{180823.1735}
\end{equation}
Here, we used that the variation of the field strength at the 
boundary is zero by the condition in \er{180727.2110}.
The right hand side of the second line shows that 
the variation of the kinetic term at the boundary 
$-\tfrac{1}{3!} \der^m (\delta C^{npq}F_{mnpq})$ is canceled 
by that of the boundary term 
$+ \tfrac{1}{3!}\der^m ( \delta C^{npq}F_{mnpq})$.
If the boundary term were not introduced, 
the variation of the Lagrangian would not vanish at the boundary.

The boundary term is also needed for the consistency 
between the energy-momentum tensor and the EOM.
If boundary term were absent, the energy-momentum tensor 
and the EOM would not be compatible with each 
other~\cite{Duncan:1989ug,Duff:1989ah}.
This can be seen as follows.
If the boundary term were absent, the Lagrangian would be written as
\begin{equation}
{\cal L}_\text{kin.,bulk} := -\df{1}{2\cdot 4!} F^{mnpq} F_{mnpq}.
\end{equation}
The energy-momentum tensor of this Lagrangian is 
\begin{equation}
 T^{mn} = 
\df{1}{3!} F^{mpqr}F^n{}_{pqr} - 
\df{1}{2\cdot 4!} \eta^{mn}F^{pqrs}F_{pqrs}
= 
- \eta^{mn} F^2 + \eta^{mn}\df{1}{2} F^2
=
 - \df{1}{2} \eta^{mn}F^2.
\end{equation}
Note that the energy-momentum tensor 
is a local quantity, and is the same whether 
boundary term is included or not
before substituting the solution of EOM of the 3-form gauge field.

The variation of the Lagrangian 
leads to the same EOM and its solution 
which is also independent of whether the boundary term is included or 
not.
Let us substitute the solution into the energy-momentum tensor and the Lagrangian.
The energy-momentum tensor is proportional to a constant:
\begin{equation}
 T^{mn} = -\df{1}{2}\eta^{mn} F^2 = -\df{1}{2}\eta^{mn} c^2.
\label{180724.2135}
\end{equation}
The on-shell Lagrangian is also merely a constant:
\begin{equation}
 {\cal L}_\text{kin.,bulk} =
-\df{1}{2\cdot 4!} \epsilon^{mnpq} \epsilon_{mnpq} c^2
= +\df{1}{2}c^2.
\label{1}
\end{equation}
We can also derive the energy-momentum tensor of the on-shell 
Lagrangian in \er{1}.
However, the Lagrangian gives the energy-momentum tensor 
which is not equal to the one in \er{180724.2135}:
\begin{equation}
 T^{mn} = +\df{1}{2}\eta^{mn} c^2.
\label{2}
\end{equation}
Therefore, if the boundary term were absent, the EOM is not 
consistent with energy-momentum tensor unless the constant $c$ 
is equal to zero
\footnote{Historically, the 3-form gauge field was used to 
consider the cosmological constant problem 
in Ref.~\cite{Hawking:1984hk}.
However, it was pointed out that the discussion led to wrong sign for the cosmological constant since the boundary term for the 3-form was not included~\cite{Duff:1989ah, Wu:2007ht}.
}. 

The boundary term resolves this problem~\cite{Duncan:1989ug,Duff:1989ah}.
We again consider the Lagrangian ${\cal L}_\text{kin.}$ in \er{3}.
In the presence of the boundary term,
the Lagrangian after substituting the solution of 
the EOM is changed as follows: 
\begin{equation}
{\cal L}_\text{kin.} = -\df{1}{2}c^2. 
\end{equation}
Then the energy-momentum tensor is given by
\begin{equation}
 T^{mn} = -\df{1}{2} \eta^{mn} c^2.
\end{equation}
Therefore, the solution of the equation of motion is consistent with
the energy-momentum tensor.

\subsection{Higher derivative term causing tachyon and/or ghost
in 3-form gauge theories}
\label{tach}
Here, we show that higher derivative terms in the form of 
derivatives on the field strength 
 in 3-form gauge theories may cause 
a tachyon as far as the canonical kinetic term
exists. 
We show that the tachyon can also be a ghost depending on parameters.

We see that
the field strength can become dynamical and tachyonic
if higher derivative terms in the form of 
derivatives of the field strength of a 3-form gauge field are present,
and if the canonical kinetic term exists.
As an example, we consider the following Lagrangian
with a term $\der^m F \der_m F$:
\begin{equation}
{\cal L}_{\der F} 
= +\df{1}{2} F^2 
+\df{\alpha}{2} \der^m F \der_m F
-\df{1}{3!} \der_m (\epsilon^{mnpq} C_{npq} F),
\label{180821.1732}
\end{equation}
where $\alpha$ is an arbitrary parameter with mass 
dimension $-2$.
The first term is the canonical kinetic term,
and the second term is a higher derivative term  which includes 
a derivative on the field strength $\der_m F$.
The third term is the corresponding boundary term.
Note that the boundary term for the second term is not needed because derivatives on the field strength at the boundary are zero by the boundary condition in \er{180727.2110}.

To see that there is a tachyonic mode,
we rewrite the Lagrangian 
by introducing new fields $F'$ and $q$:
\begin{equation}
 {\cal L}'_{\der F}
= +\df{1}{2} F'^2 
+\df{\alpha}{2} \der^m F' \der_m F'
-\df{1}{3!} \der_m (\epsilon^{mnpq} C_{npq} F')
+q \(F' - \df{1}{3!}\epsilon^{mnpq}\der_m C_{npq}\). 
\label{180821.1749}
\end{equation}
Here, $F'$ is a pseudo-scalar field independent of the 3-form gauge field.
We assume that the boundary condition for $F'$ is 
the same as $F$: $F'|_\text{bound.} = -c$.
$q$ is a Lagrange's multiplier field
whose EOM gives us the original Lagrangian in \er{180821.1732}.
Vanishing of the variation of $C_{mnp}$ at the boundary 
requires the boundary condition for $q$:
$q|_\text{bound.} = -F'|_\text{bound.} = c$. 
The EOM of $C_{mnp}$ implies that $q$ is a local constant, 
which is equal to $c$ by the boundary condition.
By substituting the solution into \er{180821.1749}, 
and redefining $F'$ as $F''= F' + c$, we obtain
\begin{equation}
 {\cal L}'_{\der F} 
= +\df{1}{2} F''^2
+\df{\alpha}{2} \der^m F'' \der_m F''
 -\df{1}{2}c^2.
\end{equation}
The first term $+\tfrac{1}{2} F''^2$ becomes a tachyon mass term.
The origin of the tachyon mass term is the canonical kinetic term with the correct sign.
Furthermore, the second term is the kinetic term for 
$F''$.
Thus, the Lagrangian in \er{180821.1749} contains 
a dynamical tachyon.
If the sign of $\alpha$ is positive,
the field $F''$ is a ghost as well.

Note that there is no tachyon, if the canonical kinetic 
term in the original Lagrangian has a wrong sign $-\tfrac{1}{2} F^2$.
Furthermore, if the parameter $\alpha$ is 
negative in addition to the kinetic term with the wrong sign,
there is neither a ghost nor a tachyon \cite{Klinkhamer:2017nfb}.

Let us make a comment on a relation to SUSY Lagrangian.
The Lagrangian in \er{180821.1732}
is obtained from the SUSY Lagrangian
in \er{180813.1621} given in Ref.~\cite{Antoniadis:2007xc}
by regarding the imaginary part 
of the auxiliary field as the field strength of the 
3-form gauge field and truncating all fields 
other than the field strength.

\subsection{Ghost/tachyon-free higher derivative 3-form gauge theories}
\label{bh}

Here, we consider a ghost/tachyon-free 
higher derivative extension 
of the 3-form gauge theory.
In the previous subsection, 
we have seen that the derivatives on the field strength 
can cause a tachyon.
Thus, we only consider a higher derivative Lagrangian 
with a function of the field strength 
but not of derivatives on the field strength.
Such higher derivative Lagrangians were previously considered in 
Refs.~\cite{Dvali:2005an,Dvali:2005zk,DAmico:2017cda}.
In this subsection, we confirm that there 
are no tachyons as well as ghosts 
in this higher derivative Lagrangian
in contrast to the previous subsection.
We determine 
the boundary term for the ghost/tachyon-free higher derivative term.
We show that the boundary terms are also needed for 
the higher derivative interactions of the 3-form gauge field.
We determine it by requiring that 
the variation of the Lagrangian at the boundary should vanish.

\subsubsection{Ghost/tachyon-free higher derivative Lagrangian: bulk part}

First of all, 
we confirm that a higher derivative term given by a function of 
the field strength is ghost/tachyon-free.
We consider the following Lagrangian
whose bulk part was  
discussed in Refs.~\cite{Dvali:2005an,Dvali:2005zk}:
\begin{equation}
{\cal L}_\text{HD}
= G(F) + {\cal L}_\text{HD,bound.}
\label{180630.1}
\end{equation}
where $G(F)$ is an arbitrary real function of $F$,
 and ${\cal L}_\text{HD,bound.}$ is the boundary term 
which we will determine below. 
One can assume that $G(F)$ includes the 
canonical kinetic term: $G(F) = +\tfrac{1}{2} F^2+\cdots$.
We impose the same boundary condition on $F$ as \ers{180725.1524}
and \eqref{180727.2110}:
\begin{equation}
 F|_{\text{bound.}} = c, 
\label{180728.0209}
\end{equation}
and
\begin{equation}
 \delta F|_{\text{bound.}} = 0. 
\end{equation}
In this Lagrangian in \er{180630.1},
we can see that there are no additional 
dynamical degrees of freedom.
The EOM of the 3-form gauge field is 
\begin{equation}
0
=
G''(F) \der^m F_{mnpq}.
\label{180630.3}
\end{equation}
The solutions of the EOM are $\der^m F_{mnpq} =0$
or $G''(F)=0$.
For $\der^m F_{mnpq} =0$, $F$ is a constant, which is 
determined by the boundary condition.
For $G''(F)=0$, the value of $F$ depends on $G(F)$,
and this value does not always satisfy the boundary condition.
Thus, we focus on the former solution $\der^m F_{mnpq} =0$.
The EOM can be solved as
\begin{equation}
 F_{mnpq} = c\epsilon_{mnpq},
\label{180717.1804}
\end{equation}
where $c$ is a constant determined by the boundary condition
in \er{180728.0209}.
For this solution, there are no additional degrees of freedom 
in the higher derivative Lagrangian since the EOM is not changed from the case of the canonical kinetic term.
Thus, there are no tachyons as well as ghosts in \er{180630.1}.

\subsubsection{Ghost/tachyon-free higher derivative Lagrangian: boundary part}

Next, we see that the boundary term is 
needed for the vanishing of the variation of the Lagrangian 
at the boundary and 
for the consistency between EOM and the 
energy-momentum tensor.
We show that 
the ghost/tachyon-free Lagrangian of the 3-form gauge field
 without the corresponding 
boundary term also gives rise to an inconsistency between the EOM and the energy-momentum tensor.

We see the inconsistency between  
the energy-momentum tensor and the EOM of the 
Lagrangian in \er{180630.1}.
On one hand, the energy-momentum tensor of 
the Lagrangian in \er{180630.1} can be calculated as
\begin{equation}
 T^{mn} = \eta^{mn}(-F G'(F)  +G(F)).
\label{180725.1925}
\end{equation}
On the other hand,
substituting the solution in \er{180717.1804} 
into the energy-momentum tensor 
in \er{180725.1925}, 
we obtain
\begin{equation}
 T^{mn} = \eta^{mn}(c G'(-c)  +G(-c)) .
\label{180728.0210} 
\end{equation}
If the boundary term ${\cal L}_\text{HD,bound.}$ 
were absent in the Lagrangian in \er{180630.1}, 
the Lagrangian would be
\begin{equation}
 {\cal L}_{\text{HD,bulk}}: = G(F),
\end{equation}
and the on-shell Lagrangian would be
\begin{equation}
{\cal L}_\text{HD,bulk} = G(-c).
\label{180728.0213} 
\end{equation}
We can see that the energy-momentum tensor which is calculated by
the on-shell Lagrangian in \er{180728.0213}, 
$T^{mn}= \eta^{mn} G(-c)$, is inconsistent with the one in 
\er{180728.0210}.

Therefore, we need the boundary term which gives us the consistent energy-momentum tensor.
We can find the boundary term by the variational principle.
The variation of the Lagrangian by the 3-form gauge field is 
\begin{equation}
\delta {\cal L}_\text{HD}  =
\df{1}{3!} G' (F) \epsilon^{mnpq}\der_m \delta C_{npq}
+\delta {\cal L}_\text{HD,bound.}.
\end{equation}
By the partial integration, we obtain the variation at the 
boundary:
\begin{equation}
\delta {\cal L}_\text{HD}=
\df{1}{3!}\der_m (\epsilon^{mnpq} G'(F) \delta C_{npq})
+\delta {\cal L}_\text{HD,bound.}
+\cdots,
\end{equation}
where the ellipsis $\cdots$ denotes
the variation in the bulk.
To cancel the variation at the boundary, 
we propose the boundary term 
corresponding to the higher derivative term:
\begin{equation}
 {\cal L}_\text{HD,bound.} 
= 
-\df{1}{3!}\der_m (\epsilon^{mnpq} G'(F) C_{npq}).
\label{180703.2001} 
\end{equation}
Note that boundary term given in \er{3} is naturally included into
the boundary term in \er{180703.2001}
by choosing $G(F) = +\tfrac{1}{2}F^2$.
To confirm whether the boundary term is consistent or not,
we consider the EOM and the energy-momentum tensor.
We start with the following Lagrangian:
\begin{equation}
 {\cal L}_\text{HD} 
= 
 {\cal L}_\text{HD, bulk} +  {\cal L}_\text{HD, bound.} 
= 
G(F)
-
\df{1}{3!}
\der_m (\epsilon^{mnpq} G'(F) C_{npq}).
\label{180725.1932}
\end{equation}
The EOM is the same as \er{180630.3},
but the on-shell Lagrangian is modified as
\begin{equation}
 {\cal L}_\text{HD}
=
G(-c)
+ c G'(-c),
\label{180703.2003} 
\end{equation}
which leads to the energy-momentum tensor which is
consistent with \er{180725.1925}:
\begin{equation}
 T^{mn} = 
\eta^{mn}(c G'(-c) +G(-c)).
\label{180725.1931}
\end{equation}
Therefore, the Lagrangian which includes the boundary term 
in \er{180725.1932} gives us the consistent energy-momentum tensor.

\section{SUSY 3-form gauge theories}
\label{sch}

In this section, we give the most general ghost/tachyon-free SUSY Lagrangian of 
a 3-from gauge field of an arbitrary order of the field strength.
In Sec.~\ref{sc}, we review a formulation of a Lagrangian 
with the quadratic kinetic term in SUSY 3-form gauge theories.
We also review a boundary term for the quadratic kinetic term.
In Sec.~\ref{sg} 
we give an example of  
SUSY higher derivative Lagrangian of 3-form gauge field containing a ghost as well as tachyon.
In Sec.~\ref{sh}, we give the most general
ghost/tachyon-free Lagrangian and its boundary term.

In this section, we use superspace to formulate manifestly SUSY theories.
Superspace is spanned by 
spacetime coordinate $(x^m)$ and fermionic coordinate
given by Grassmann variables $(\theta^\alpha,\b\theta_{\d\alpha})$.
Here, undotted and dotted Greek letters $\alpha, \beta,...$ and 
$\d\alpha,\d\beta,...$ denote undotted and dotted spinors, 
respectively.

\subsection{SUSY 3-form gauge theory with canonical kinetic term}
\label{sc}

In this subsection, we review formulations of 
Lagrangians with quadratic kinetic terms
in SUSY 3-form gauge theories.
This subsection is essentially based on 
Refs.~\cite{Gates:1980ay,Groh:2012tf,Farakos:2017jme}.

\subsubsection{3-form prepotential}
Here we explain how to embed a 3-form gauge field into a superfield.
In superspace, fields are embedded into superfields.
A 3-form gauge field $C_{mnp}$ is embedded into a real superfield $X$
($X^\dg = X$)~\cite{Gates:1980ay}:
\begin{equation}
 C_{mnp} = \df{\sr{2}}{8}\epsilon_{mnpq} (\b\sigma^q)^{\d\alpha\alpha}
[D_\alpha, \b{D}_{\d\alpha}] X|.
\end{equation}
Here, the vertical bar ``$|$'' denotes 
$\theta= \b\theta =0$ projection in superspace.
The derivatives $D_\alpha$ and $\b{D}_{\d\alpha}$
are SUSY covariant spinor derivatives.
Following Ref.~\cite{Gates:1983nr}, 
we call $X$ ``3-form prepotential'' in this paper.
An infinitesimal superfield 
gauge transformation of the 3-form prepotential is 
given by
\begin{equation}
\delta_\text{3,SUSY} X 
 =
 \df{1}{2i}(D^\alpha \Upsilon_\alpha 
- \b{D}_{\d\alpha}\b\Upsilon^{\d\alpha}),
\end{equation}
where $\Upsilon_\alpha$ is a chiral superfield $\b{D}_{\d\alpha}\Upsilon_{\alpha}=0$.
Here, $\delta_\text{3,SUSY}$ 
denotes the infinitesimal gauge transformation 
of the 3-form prepotential.
Since $L :=  \tfrac{1}{2i}(D^\alpha \Upsilon_\alpha 
- \b{D}_{\d\alpha}\b\Upsilon^{\d\alpha})$
is a real linear superfield 
satisfying $D^\alpha D_\alpha L = 
\b{D}_{\d\alpha} \b{D}^{\d\alpha} L=0$,
 the gauge transformation can be simply rewritten by the 
real linear superfield as
\begin{equation}
\delta_\text{3,SUSY} X  = L.
\end{equation}
The gauge transformation of the bosonic gauge field is 
included in this superfield gauge transformation:
\begin{equation}
\begin{split}
 \delta_\text{3,SUSY} C_{mnp} 
&
= 
\df{\sr{2}}{8}\epsilon_{mnpq} (\b\sigma^q)^{\d\alpha\alpha}
[D_\alpha, \b{D}_{\d\alpha}] \delta_\text{3,SUSY} X|
\\
&
=  
 \df{\sr{2}}{8}\epsilon_{mnpq} (\b\sigma^q)^{\d\alpha\alpha}
[D_\alpha, \b{D}_{\d\alpha}] 
 \df{1}{2i}(D^\gamma \Upsilon_\gamma 
- \b{D}_{\d\gamma}\b\Upsilon^{\d\gamma})|
\\
&
=\der_m \lambda_{np}
+
\der_n \lambda_{pm}
+
\der_p \lambda_{mn},
\end{split}
\end{equation}
where the gauge parameter $\lambda_{mn}$ is embedded into 
the chiral superfield $\Upsilon_\alpha$ as
\begin{equation}
 \lambda_{mn}
= \df{\sr{2}}{2i} 
\((\sigma_{mn})_\alpha{}^\beta D^\alpha \Upsilon_\beta 
-(\b\sigma_{mn})^{\d\alpha}{}_{\d\beta} 
\b{D}_{\d\alpha} \b\Upsilon^{\d\beta}\)|.
\end{equation}
Note that a prepotential for a 1-form (vector superfield) is 
also a real superfield, 
but the gauge transformation law of the 1-form prepotential
 is different from that of the 3-form prepotential.

The field strength of the 3-form gauge field
is embedded into a chiral superfield $Y$ (and its Hermitian conjugate),
which is given by the 3-form prepotential as follows:
\begin{equation}
 Y: = -\df{1}{4} \b{D}^2 X.
\end{equation}
Here, the second order spinor derivative 
$\b{D}^2 := \b{D}_{\d\alpha} \b{D}^{\d\alpha}$ 
acts on $X$ as a chiral projection $\b{D}_{\d\alpha }\b{D}^2= 0$.
Therefore, $Y$ is a chiral superfield: $\b{D}_{\d\alpha} Y=0$.
We will call $Y$ ``4-form field strength superfield'' in this paper.
Note that $Y$ is gauge invariant: 
$\delta_\text{3,SUSY} Y = -\tfrac{1}{4}\b{D}^2 \delta_\text{3,SUSY} X
 = -\tfrac{1}{4}\b{D}^2 L=0$.
The bosonic field strength $F_{mnpq}$ is embedded into 
the imaginary part of the auxiliary field of $Y$:
\begin{equation}
 F_{mnpq} = \df{\sr{2} i}{8}\epsilon_{mnpq} (D^2 Y - \b{D}^2 \b{Y})|,
\end{equation}
where $D^2 := D^\alpha D_\alpha$.
This is equivalently written by using $F$ as
\begin{equation}
 F = -\df{\sr{2} i}{8} (D^2 Y - \b{D}^2 \b{Y})|.
\end{equation}

In SUSY theories, the 3-form gauge field has dynamical superpartners.
Since the 3-form gauge field has one off-shell degree of freedom,
there are also off-shell fermionic degrees of freedom as well as 
additional bosonic ones.
The superpartners of the 3-form gauge field can be found as components of the chiral superfield $Y$.
They are one complex scalar $y$ (two bosons), 
one Weyl fermion $\chi_\alpha$ (four fermions), and one real auxiliary scalar field $H$ (one boson).
We define the components as follows:
\begin{equation}
 y:= Y|, 
\quad
\chi_\alpha := \df{1}{\sr{2}} D_\alpha Y|,
\quad
H := - \df{\sr{2}}{8} (D^2 Y + \b{D}^2 \b{Y})|.
\end{equation}
Note that the dynamical degrees of freedom are 
$y$ and two of $\chi_\alpha$,
thus the on-shell bosonic degrees of freedom are also equal to 
the fermionic ones.
It is convenient to define the following 
complex auxiliary field ${\cal F}$:
\begin{equation}
 {\cal F}
:=-\df{1}{4}D^2 Y|
= \df{1}{\sr{2}}(H- iF).
\end{equation}

We impose the following boundary conditions for the component fields.
One is the boundary condition for $F$, which is the 
same as the previous one:
\begin{equation}
F|_\text{bound.}= -c.
 \label{180825.1358}
\end{equation}
Other boundary conditions are imposed so that 
SUSY and gauge invariance are preserved at the boundary~\cite{Groh:2012tf}:
\begin{equation}
 \delta y |_\text{bound.} = 0, 
\quad
 \delta \chi_\alpha |_\text{bound.}
 = 0, 
\quad
 \delta H |_\text{bound.}
 = 0, 
\qtq{and}
 \delta F |_\text{bound.}
 = 0.
\label{180815.1853}
\end{equation}
Here, $\delta$ denotes variations of fields.
In the superfield language, the boundary conditions can be 
written as
\begin{equation}
 \b{D}^2 \delta X ||_\text{bound.}=0,
\quad
D_\alpha \b{D}^2 \delta X ||_\text{bound.}=0,
\quad
D^2 \b{D}^2 \delta X ||_\text{bound.}=0,
\end{equation}
and their Hermitian conjugates.
Here, $||_\text{bound.}$ denotes the 
$\theta =\b\theta =0$ projection at the boundary. 
The other boundary conditions for 
the higher order spinor derivatives on 
$\b{D}^2 \delta X$ and $D^2 \delta X$
 are also assumed to be zero.
The examples used later are 
\begin{equation}
[D_\alpha,\b{D}_{\d\alpha }] \b{D}^2 \delta X ||_\text{bound.}=0,
\quad
\b{D}^2 D_\alpha \b{D}^2 \delta X ||_\text{bound.}=0,
\label{180910.0412}
\end{equation}
and so on.
Meanwhile, as in the non-SUSY case, 
we do not impose a specific boundary condition 
for the gauge field $C_{mnp}$, since 
the gauge field is not gauge invariant.

\subsubsection{SUSY Lagrangian with canonical kinetic term: bulk part}
The Lagrangian with a canonical kinetic term is given by 
\begin{equation}
\begin{split}
  {\cal L}_\text{kin.,SUSY} 
&= 
-\df{1}{8}\int d^2 \theta \b{D}^2 Y\b{Y}
-\df{1}{8}\int d^2 \b\theta D^2 Y\b{Y}
+{\cal L}_\text{kin.,SUSY,bound.}
\\
&
=
-\der^m y\der_m \b{y}
-i \b\chi_{\d\alpha} (\b\sigma^m)^{\d\alpha\alpha}\der_m \chi_\alpha
+\df{1}{2}H^2 
-\df{1}{2\cdot 4!} F^{mnpq} F_{mnpq} 
+{\cal L}_\text{kin.,SUSY,bound.}
\end{split}
\label{180825.1329}
\end{equation}
Here $\int d^2\theta = -\tfrac{1}{4}D^2 |$ 
is the F-type integration.
We use 
$-\tfrac{1}{8}(\int d^2\theta \b{D}^2 + \int d^2\b\theta D^2) $ 
for the D-type integration instead of the 
conventional $\int d^4\theta $
in order to fix the definition of the D-type integration.
Our choice for the D-type integration may have a natural extension 
to Poincar\'e and conformal 
SUGRA~\cite{Cremmer:1982en,Kugo:1982cu,Kugo:1982mr,Wess:1992cp,Butter:2009cp,Kugo:2016zzf,Kugo:2016lum}.
Note that we have neglected the boundary terms which do not depend on the field strength $F_{mnpq}$. 
Such boundary terms are not relevant to our discussion.
As in the non-SUSY case, 
we derive the EOM for the 3-form gauge field in the SUSY case.
This can be evaluated by the variation of the 3-form prepotential
 in the bulk:
\begin{equation}
 0 = -\df{1}{4}(\b{D}^2 \b{Y} + D^2Y).
\label{180825.1344}
\end{equation}
The $\theta = \b\theta = 0$ 
component of the EOM leads to the EOM for the 
real auxiliary field $H$:
\begin{equation}
 H=0.
\end{equation}
The EOM for $F$
 can be found in the $[D_\alpha,\b{D}_{\d\alpha}]$ 
component of the EOM in \er{180825.1344}:
 \begin{equation}
 0 = -\df{1}{4}[D_\alpha,\b{D}_{\d\alpha} ]
(\b{D}^2 \b{Y} + D^2 Y)|
= -\df{i}{2}(\sigma^m)_{\alpha\d\alpha}
\der_m 
(\b{D}^2 \b{Y} - D^2 Y)|
 = 
2\sr{2} (\sigma^m)_{\alpha\d\alpha}
\der_m F,
\label{180825.1357}
\end{equation}
where we have used 
 $[D_\alpha,\b{D}_{\d\alpha}](\text{chiral})
= +2i\der_{\alpha\d\alpha} (\text{chiral})$ and 
$[D_\alpha,\b{D}_{\d\alpha}](\text{anti-chiral})
= -2i\der_{\alpha\d\alpha} (\text{anti-chiral})$.
Here, $\der_{\alpha\d\alpha}$ is the spinor representation of 
the spacetime derivative: 
$\der_{\alpha\d\alpha} = (\sigma^m)_{\alpha\d\alpha}\der_m$.
Thus, $F$ is equal to a constant, which is determined by 
the boundary condition for $F$ in \er{180825.1358}:
\begin{equation}
 F = -c.
\end{equation}
We can rewrite the solution to the EOM in terms of the 
complex auxiliary field ${\cal F}$ as
\begin{equation}
 {\cal F} = -\df{1}{4}D^2 Y| = i\df{c}{\sr{2}}.
\end{equation}

Since the 4-form field strength superfield is described by the 
chiral superfield, the kinetic term $Y\b{Y}$ can be generalized to 
a K\"ahler potential $K(Y, \b{Y})$.
Further, the 4-form field strength superfield 
can have a superpotential.
Therefore, the Lagrangian with quadratic derivative terms
is given by
\begin{equation}
{\cal L}_{KW}
 = 
\(
-\df{1}{8} \int d^2\theta \b{D}^2 
K(Y,\b{Y}) 
+\int d^2\theta W(Y) 
+\hc 
\)
+{\cal L}_{KW,\rm bound.}
\label{180718.1611}
\end{equation}
Here, ${\cal L}_{KW,\rm bound.}$ is the boundary term 
for the above Lagrangian, which we will determine below.
The EOM for the field strength can be obtained in the same way as the case of the Lagrangian in \er{180825.1329}.
\subsubsection{SUSY Lagrangian with canonical kinetic term: boundary part}

In the variation of the Lagrangian in \er{180825.1329}
by the 3-form prepotential, 
we have neglected the variation at the boundary, which we should discuss carefully.
Therefore, we consider the boundary term for 
the canonical kinetic term~\cite{Groh:2012tf,Farakos:2017jme}.
The term ${\cal L}_\text{kin.,SUSY,bound.}$ in \er{180825.1329} 
is a boundary term corresponding to the 
canonical kinetic term.
As in the non-SUSY case, 
 there is generally a non-trivial boundary condition for 
the field strength of the 3-form.
The boundary term is obtained by either the variational 
principle~\cite{Groh:2012tf} or a dual 
formulation~\cite{Farakos:2017jme}.
Since we can straightforwardly obtain 
the boundary term by the variational principle in terms of 
the 3-form prepotential only, we use the former option here.
The latter option is summarized in Appendix~\ref{sd}.

In the following discussion, we consider
the variation of the Lagrangian by the 3-form 
prepotential, and find the term proportional to 
$\der_m \delta C_{npq}$, which gives a nontrivial variation at the boundary as in the non-SUSY case in Sec.~\ref{bc}.
We then introduce a boundary term which cancels the variation at the boundary.

The variation of the kinetic term in \er{180825.1329} by 
the 3-form prepotential is
\begin{equation}
\begin{split}
\delta {\cal L}_\text{kin., SUSY}
=
\df{1}{32}
\Bigg(
&\int d^2 \theta \b{D}^2 (\b{Y} \b{D}^2 \delta X)
+\int d^2 \theta \b{D}^2 (Y D^2 \delta X)
\\
&\quad
+\int d^2 \b\theta D^2 (\b{Y} \b{D}^2 \delta X) 
+\int d^2 \b\theta D^2 ( Y D^2 \delta X)
\Bigg) 
+
\delta {\cal L}_\text{kin.,SUSY,bound.}.
\end{split}
\label{180717.2004}
\end{equation} 
Here, we will show that the second and the third terms 
in the right hand side
are equal to the fourth and the first terms, respectively.
For example, we consider the third term 
$\int d^2 \b\theta D^2 (\b{Y} \b{D}^2 \delta X) $.
By using the identity 
\begin{equation}
 D^2 \b{D}^2 - \b{D}^2 D^2 
= 
-4i \der^{\d\alpha\alpha}[D_\alpha, \b{D}_{\d\alpha}],
\label{180717.2005}
\end{equation}
 this term is equal to
\begin{equation}
\begin{split}
 \int d^2 \b\theta D^2 (\b{Y} \b{D}^2 \delta X) 
&=
-\df{1}{4} \b{D}^2 D^2 (\b{Y} \b{D}^2 \delta X) |
\\
&=
-\df{1}{4}  D^2 \b{D}^2 (\b{Y} \b{D}^2 \delta X) |
-i \der^{\d\alpha\alpha}[D_\alpha ,\b{D}_{\d\alpha}] 
(\b{Y} \b{D}^2 \delta X) |.
\end{split}
\end{equation}
The difference between the first and the third 
terms in \er{180717.2004} is the total derivative term
$ -i \der^{\d\alpha\alpha}([D_\alpha ,\b{D}_{\d\alpha}] 
(\b{Y} \b{D}^2 \delta X)|) $.
The total derivative is equal to zero because
the variation  
$\delta Y|$, $D_\alpha \delta Y|$, and $\der_m \delta Y |$ 
are assumed to be zero in \er{180815.1853} 
 to preserve SUSY at the 
boundary.
Therefore, \er{180717.2004} can be rewritten as 
\begin{equation}
\delta {\cal L}_\text{kin., SUSY}
=
\df{1}{16}
\(
\int d^2 \theta \b{D}^2 (\b{Y} \b{D}^2 \delta X)
+\int d^2 \b\theta D^2  (Y D^2 \delta X)
\)
+\delta {\cal L}_\text{kin.,SUSY,bound.}
.
\label{180717.2023}
\end{equation} 
By using $D_\alpha D^2 =0$, we can further rewrite the variation as
\begin{equation}
\delta {\cal L}_\text{kin., SUSY}
=
\df{1}{16}
\(
\int d^2 \theta (\b{D}^2 \b{Y}) \b{D}^2 \delta X
+\int d^2 \b\theta (D^2  Y) D^2 \delta X
\)
+\delta {\cal L}_\text{kin.,SUSY,bound.}
.
\label{180717.2024}
\end{equation} 
For later use, we define a chiral superfield 
\begin{equation}
 T:= -\df{1}{4}\b{D}^2 \b{Y},
\label{180909.0209}
\end{equation}
whose $\theta = \b\theta =0$ component is 
\begin{equation}
  T| 
= -\df{1}{4}\b{D}^2 \b{Y}|
 = \df{1}{\sr{2}}(H + iF).
\end{equation}
Using the chiral superfield, the variation is manifestly written 
by the product of the chiral superfield $T$ and 
$\b{D}^2 \delta X$.
This structure will be useful in the higher derivative case
 discussed later.

In this variation of the Lagrangian, we show that there is a term proportional to $\der_m \delta C_{npq}$ which gives rise to the variation of the 3-form at the boundary.
This can be seen by the component expansion:
\begin{equation}
\begin{split}
\delta {\cal L}_\text{kin., SUSY}
&=
-\df{1}{4}
\(
\int d^2 \theta T \b{D}^2 \delta X
+\int d^2 \b\theta \b{T} D^2 \delta X
\)
+\delta {\cal L}_\text{kin.,SUSY,bound.}
\\
&= 
\df{1}{16} iT_I (D^2\b{D}^2 -\b{D}^2 D^2) \delta X|
+\delta {\cal L}_\text{kin.,SUSY,bound.}
+\cdots, 
\end{split}
\label{180717.2025}
\end{equation}
where $T_I = \tfrac{1}{2i}(T-\b{T})$ is the imaginary part of $T$, 
and the ellipsis $\cdots$ means the terms which are not 
related to the variation of the 3-form.
The first term can be calculated as
\begin{equation}
\df{1}{16} i T_I (D^2\b{D}^2 -\b{D}^2 D^2) \delta X|
 = 
\df{1}{16} iT_I (-4i) \der^{\d\alpha \alpha} [D_\alpha,\b{D}_{\d\alpha}]
\delta X |
= \df{1}{3!}
F\epsilon^{mnpq} \der_m \delta C_{npq}.
\label{180718.1533}
\end{equation}
Therefore, the variation gives us the following boundary term
\begin{equation}
\df{1}{16} \der^{\d\alpha \alpha}(iT_I (-4i)  [D_\alpha,\b{D}_{\d\alpha}]
\delta X) |
= \df{1}{3!} \der_m(F\epsilon^{mnpq} \delta C_{npq}).
\end{equation}
To cancel the variation at the boundary, 
we determine the boundary term 
${\cal L}_\text{kin.,SUSY,bound.}$ as follows:
\begin{equation}
{\cal L}_\text{kin.,SUSY,bound.}
=
\df{i}{4}
\(\int d^2\theta \b{D}^2 - \int d^2 \b\theta D^2\)
T_I X.
\label{180718.1541}
\end{equation}
In fact, 
this Lagrangian is a boundary term which cancels \er{180718.1533}
 since the identity of the spinor derivatives gives us 
\begin{equation}
 \df{i}{4}
\(\int d^2\theta \b{D}^2 - \int d^2 D^2\)
T_I X
=
-\df{i}{16}(D^2\b{D}^2 - \b{D}^2 D^2) T_I X|
= -\df{1}{4} \der^{\d\alpha\alpha} [D_\alpha,\b{D}_{\d\alpha}] T_I X|.
\end{equation}
In the Wess--Zumino (WZ) gauge~\cite{Gates:1983nr} where
\begin{equation}
X|= D_\alpha X| = \b{D}_{\d\alpha} X|=0, 
\label{180910.0422}
\end{equation}
 the boundary term is
\begin{equation}
{\cal L}_\text{kin.,SUSY,bound.}
=
-\df{1}{3!}\der_m 
(F \epsilon^{mnpq} C_{npq}).
\end{equation}
Thus, the Lagrangian in \er{180718.1541} gives correct boundary term.

For convenience,
we 
refer to
the following $\theta$-integration 
given in Ref.~\cite{Farakos:2017jme}
\begin{equation}
 -\df{1}{4}\cdot \df{1}{2i}
\(
\int d^2\theta \b{D}^2 - \int d^2\b\theta D^2 
\)
\end{equation}
as an I(maginary)-type integration in this paper.
This integration is convenient to 
describe boundary terms such as the term in \er{180718.1541}.

Of course, the boundary term is also needed in the case
where the Lagrangian is written by a K\"ahler potential 
and a superpotential in \er{180718.1611}.
The boundary term is found in the same way as the previous 
case.
The difference is merely the choice of the chiral superfield $T$.
We will see the boundary term more precisely.
The variation of the Lagrangian in \er{180718.1611} by 
the 3-form gauge field is 
\begin{equation}
\begin{split}
 \delta {\cal L}_{KW}
& = \df{1}{32}\(\int d^2\theta \b{D}^2 + \int d^2\b\theta D^2\)
\(\pd{K}{Y} \b{D}^2 \delta X + \pd{K}{\b{Y}} D^2 \delta X 
\)
\\
&
\quad
-\df{1}{4}\int d^2\theta \pd{W}{Y} \b{D}^2 \delta X
 -\df{1}{4}\int d^2\theta \pd{\b{W}}{\b{Y}} D^2 \delta X
+
\delta {\cal L}_{KW,\rm bound.}.  
\end{split}
\label{180906.1714}
\end{equation}
We can further rewrite the variation as
\begin{equation}
\begin{split}
 \delta {\cal L}_{KW}
& = -\df{1}{4}
\int d^2\theta 
\(
-\df{1}{4}\b{D}^2 \pd{K}{Y} + \pd{W}{Y}
\)
 \b{D}^2 \delta X 
-\df{1}{4} \int d^2\b\theta 
\(
-\df{1}{4} D^2 \pd{K}{\b{Y}}
+\pd{\b{W}}{\b{Y}}
\) D^2 \delta X
\\
&
\quad
+\delta{\cal L}_{KW,\rm bound.}
\end{split}
\label{180718.2002}
\end{equation}
because the following equation holds from the boundary conditions
 in \ers{180815.1853} and \eqref{180910.0412}:
\begin{equation}
 \der^{\d\alpha \alpha}[D_\alpha,\b{D}_{\d\alpha}] 
\pd{K}{Y} \b{D}^2 \delta X|=0.
\end{equation}
If we define the chiral superfield $T_{KW}$
as
\begin{equation}
 T_{KW}
:=
-\df{1}{4}\b{D}^2 \pd{K}{Y} + \pd{W}{Y}, 
\label{180909.0215}
\end{equation}
the variation can be simply written as 
\begin{equation}
\begin{split}
 \delta {\cal L}_{KW}
& = -\df{1}{4}
\int d^2\theta T_{KW} \b{D}^2 \delta X 
-\df{1}{4} \int d^2\b\theta 
 \b{T}_{KW} D^2 \delta X
+\delta {\cal L}_{KW,\rm bound.}
.
\end{split}
\label{180718.2003}
\end{equation}
Since the variation has the same structure as \er{180717.2023},
we can repeat the same procedure as the previous case.
The variation at the boundary is 
\begin{equation}
\begin{split}
 \delta {\cal L}_{KW}|_{\text{bound.}}
& = \df{1}{16}(D^2\b{D}^2 -\b{D}^2 D^2)  i T_{I, KW} \delta X
+{\cal L}_{KW,\rm bound.}
\\
&
= \df{1}{4}\der^{\d\alpha \alpha }  [D_\alpha,\b{D}_{\d\alpha}]
T_{I, KW} \delta X
+\delta {\cal L}_{KW,\rm bound.},
\end{split}
\label{180718.2004}
\end{equation}
where $T_{I, KW} = \tfrac{1}{2i} (T_{KW}- \b{T}_{KW})$
is the imaginary part of $T_{KW}$.
Therefore, we introduce the following boundary term by using 
the I-type integration as follows:
\begin{equation}
 {\cal L}_{KW,\text{bound.}}
=
\df{i}{4}
\(\int d^2\theta \b{D}^2 - \int d^2 \b\theta D^2\)
T_{I,KW} X.
\label{180718.2021}
\end{equation}
Thus, we have determined the boundary term for the Lagrangian
with the K\"ahler potential and the superpotential.

\subsection{Higher derivative term causing ghost and tachyon in SUSY
3-form gauge theories}
\label{sg}
In this subsection, we consider 
higher derivative Lagrangians which may cause 
tachyons as well as ghosts in SUSY 3-form gauge 
theories.
The following discussion is an extension of the discussion in 
Ref.~\cite{Antoniadis:2007xc} for higher derivative chiral superfields.

We can expect that higher derivative Lagrangians with arbitrary order of the auxiliary field can be obtained by choosing an arbitrary function of $D^2 Y$ and its Hermitian conjugate.
However, a naively constructed higher derivative
Lagrangian gives rise to a ghost.

We will explain this more concretely.
For example, we may naively consider the following 
higher derivative Lagrangian~\cite{Antoniadis:2007xc} 
\begin{equation}
 {\cal L}_{\der F,\rm SUSY}
 = 
-\df{1}{8}
\(\int d^2\theta \b{D}^2 Y \b{Y}
+
\df{\alpha}{16} \int d^2 \theta 
\b{D^2}(\b{D}^2 \b{Y} D^2 Y)
+\hc
\)
+{\cal L}_{\der F,\rm SUSY,bound.},
\label{180813.1621}
\end{equation}
where 
${\cal L}_{\der F,\rm SUSY,bound.}$ is the boundary term 
for 
${\cal L}_{\der F,\rm SUSY}$, 
which is not relevant in this discussion.
We will show that the Lagrangian contains an Ostrogradsky's ghost
as well as a tachyon.
We can see these unstable modes by the component expression. 
The bosonic sector of the component Lagrangian is
\begin{equation}
 {\cal L}_{\der F,\rm SUSY}
 = 
-\der^m y \der_m \b{y} 
+|{\cal F}|^2
+
\alpha
(|\square y|^2 
-\der^m {\cal F}\der_m \b{\cal F}
).
\label{180903.1729}
\end{equation}
Here, we have abbreviated the boundary term.
We briefly show that there is a ghost
from the higher derivative term $|\square y|^2$. 
This can be seen by an auxiliary field method.
The higher derivative term $|\square y|^2$ can be rewritten 
by adding a new complex scalar field (auxiliary field) $\psi$ as
\begin{equation}
 {\cal L}'_{\der F,\rm SUSY}
 = 
-\der^m y \der_m \b{y} 
+|{\cal F}|^2
+
\alpha
(-|\psi|^2 
-\der^m \psi \der_m  \b{y}-  \der^m \b\psi \der_m y
-\der^m {\cal F}\der_m \b{\cal F}
).
\label{180903.1858}
\end{equation}
The EOM for $\psi$ gives us the original Lagrangian in \er{180903.1729}.
Instead, we 
obtain one negative eigenvalue by diagonalizing quadratic derivative terms.
The eigenmode of the negative eigenvalue corresponds to a ghost.

We also show that there are also dynamical tachyons.
The tachyons are 
the real auxiliary field $H$ and the field strength $F$.
These fields are now dynamical because of the higher derivative term
$-\alpha \der^m {\cal F}\der_m \b{\cal F} 
= -\tfrac{\alpha}{2} (\der^m H\der_m H + \der^m F\der_m F)$.
The terms are tachyonic because the term 
$|{\cal F}|^2 = \tfrac{1}{2} (H^2 +F^2)$
in \er{180903.1858} 
becomes mass terms with wrong signs.
The origin of the wrong signs are the canonical kinetic term,
but not the sign of the coefficient of the higher derivative term $\alpha$.
Therefore, the tachyons exist
as far as 
we include the canonical kinetic term.
The presence of the tachyons has the same structure as 
the bosonic model in Sec.~\ref{tach}.

The ghosts and tachyons can be more simply seen by a SUSY auxiliary 
method~\cite{Antoniadis:2007xc} than the above discussion.
In superspace, we can rewrite the Lagrangian in \er{180813.1621}
by adding $\Psi$ and $\Sigma$ as
\begin{equation}
\begin{split}
& {\cal L}'_{\der F,\rm SUSY}
\\
& = 
\(
-\df{1}{8}
\int d^2\theta \b{D}^2 Y \b{Y}
-\df{\alpha}{8}
\int d^2\theta \b{D}^2 \Psi\b\Psi
+
 \int d^2 \theta \Sigma \(\Psi + \df{1}{4} \b{D}^2 \b{Y}\)+\hc
\)
\\
&\quad
+
 {\cal L}'_{\der F,\rm SUSY,bound.}
\\
&
=
\(
-\df{1}{8}
\int d^2 \theta \b{D}^2
(|Y- \Sigma|^2  - \Sigma \b{\Sigma})
-\df{\alpha}{8}
 \int d^2 \theta \b{D}^2 \Psi \b\Psi
+
 \int d^2 \theta \Sigma \Psi  +\hc \)
\\
&\quad
+ {\cal L}'_{\der F,\rm SUSY,bound.}.
\end{split}
\label{180903.1952}
\end{equation}
Thus, dynamical superfields are $Y- \Sigma$, $\Sigma$ and $\Psi$.
By seeing the sign, $\Sigma$ becomes a ghost.
Furthermore, $\Psi$ becomes a tachyon.
This can be seen by solving EOM for the auxiliary fields of 
$\Sigma$ and $\Psi$, the on-shell scalar potential $V$ is 
\begin{equation}
 V = - |\psi|^2 + \df{1}{\alpha} |\sigma|^2.
\end{equation}
Here, $\psi = \Psi|$ and $\sigma = \Sigma|$.
Thus, the chiral superfield $\Sigma$ can be regarded as
an independent dynamical field.
Therefore, we cannot use the Lagrangian in \er{180813.1621} 
to construct ghost/tachyon-free Lagrangians.

One may wonder whether there are 
also a ghost or tachyons in \er{180813.1621} 
or not when we put $\alpha=0$. 
Of course, there should not be such a ghost or tachyons.
This can be seen as follows.
We can use the above auxiliary method even in the case that 
higher derivative terms are absent $\alpha=0$.
The EOM of $\Psi$ leads to $\Sigma=0$ if $\alpha=0$.
Thus, $\Sigma$ is not a dynamical field.
Furthermore, $\Psi$ drops out of the Lagrangian.
Therefore, there is 
 neither a ghost nor a tachyon 
in this case as expected.

We make a comment on the sign of the kinetic term
in \er{180813.1621}.
In Sec.~\ref{tach}, we have mentioned the sign of the 
kinetic term.
If the canonical kinetic term of the field strength 
has a wrong sign, there is no tachyon.
However, the kinetic term of $y$ has also the wrong sign
by SUSY.
Therefore, $y$ becomes a ghost in this case.

\subsection{SUSY ghost/tachyon-free higher derivative 3-form gauge theories}
\label{sh}

In this subsection, we give the most general ghost/tachyon-free Lagrangian of SUSY 3-form gauge theories.
First, we present a ghost/tachyon-free Lagrangian at an arbitrary order 
of the field strength, and then we specify the boundary term 
for the ghost/tachyon-free Lagrangian.

\subsubsection{SUSY ghost/tachyon-free higher derivative 
Lagrangian: bulk part}

The previously known ghost/tachyon-free Lagrangian of 3-form gauge theories 
is at most the fourth order of the auxiliary field, 
{\it i.~e.}~the fourth order of the field strength of the 3-form gauge field.
Here, we construct the most general ghost/tachyon-free higher derivative Lagrangian
of an arbitrary order of the field strength.
To this end,
 we use the fact that the field strength of the 3-form gauge field is embedded into a chiral superfield $Y$.
Therefore, ghost/tachyon-free Lagrangian for the SUSY 3-form gauge field 
can be formulated by that of the chiral 
superfield~\cite{Khoury:2010gb}.

We present the following Lagrangian:
\begin{equation}
\begin{split}
& {\cal L}_\text{HD,SUSY}
\\
&
= 
-\df{1}{8\cdot 16}
\(
\int d^2 \theta \b{D}^2 \Lambda(Y,\b{Y}, \der_m Y, \der_m \b{Y},
D^2 Y, \b{D^2}\b{Y}) (D^\alpha Y)(D_\alpha Y)
(\b{D}_{\d\alpha} \b{Y}) (\b{D}^{\d\alpha} \b{Y})
+\hc
\)
\\
&
\quad
+
{\cal L}_\text{HD,SUSY,bound.}
\end{split}
\label{gf3}
\end{equation}
Here, ${\cal L}_\text{HD,SUSY,bound.}$
is the boundary term, which we will discuss later.
This Lagrangian is a natural extension of ghost-free Lagrangian of 
a chiral superfield~\cite{Khoury:2010gb,Nitta:2014pwa}.
Previously known ghost-free Lagrangian  
is that $\Lambda$ is a function of $Y$, $\b{Y}$ \cite{Khoury:2010gb}, 
$\der_m Y$, and $\der_m \b{Y}$ \cite{Nitta:2014pwa}.
One new point is that
$\Lambda$  in Eq.~(\ref{gf3}) can also be a function of 
 $D^2 Y$ and $\b{D}^2 \b{Y}$. 
The Lagrangian presented in Ref.~\cite{Ciupke:2016agp} is more relaxed
and consequently contains an Ostrogradsky's ghost. 
This can be shown by the auxiliary method in Ref.~\cite{Dudas:2015vka}.

We prove that the Lagrangian in \er{gf3} is ghost/tachyon-free.
One way to show it 
is to calculate the component expression of the Lagrangian.
Another way is using the auxiliary method, which we 
summarize in Appendix~\ref{sgf}.
Here, we use the former way.
We show that there are no ghosts/tachyons in the purely bosonic sector.
If there are no ghosts in the bosonic sector, 
there should be no ghosts/tachyons in the fermionic sector because of SUSY. 

The bosonic sector of the component Lagrangian is 
\begin{equation}
\begin{split}
& {\cal L}_\text{HD,SUSY,boson}
\\
&
=
\Lambda (y, \b{y},\der_m y, \der_m \b{y}, -2\sr{2}(H-iF), -2\sr{2}(H+iF))
\\
&\qquad
\times 
\Big(\df{1}{4}(H^2 + F^2)^2 - \der^n y \der_n \b{y} (H^2 + F^2)
+ (\der^n y \der_n y)(\der^p \b{y} \der_p \b{y}) \Big)
+{\cal L}_\text{HD,SUSY,bound.}.
\end{split}
\label{180813.2157}
\end{equation}
Since there are no $\der \der y$
 terms in the bosonic sector of the Lagrangian, 
we conclude that
there are no ghosts in the bosonic sector.
Because there are no $\der F$ terms,
the field strength is not dynamical, 
and there is also no tachyonic mode 
which is discussed in Sec.~\ref{tach} and \ref{sg}.
By SUSY transformation, we conclude that there should be no fermionic ghosts in the Lagrangian as well.

Some comments are in order.
One comment is that 
the ghost/tachyon-free Lagrangian in \er{gf3} 
can be extended into
a system with matter chiral superfield (ordinary 
chiral superfield).
That is, we can relax the assumption that 
a chiral superfield $Y$ is related to a real superfield 
$Y = -\tfrac{1}{4} \b{D}^2 X$.
In this case,  we generally do not need the boundary terms which we will discuss later for the 3-form gauge theories.

Another comment is that we can straightforwardly extend 
the ghost/tachyon-free Lagrangian into a case of 
multicomponent chiral superfields $\Phi^i$ ($i = 1,...,n$) 
 with couplings of a K\"ahler potential $K(\Phi^i,\b\Phi^{i^*})$
and a superpotential $W(\Phi^i)$:
\begin{equation}
\begin{split}
{\cal L}_\text{HD,multi}
&=
-\df{1}{8}
\int d^2\theta \b{D}^2 K(\Phi^i,\b\Phi^{i^*})
+\int d^2 \theta W(\Phi^i)
\\
&\quad
-\df{1}{8\cdot 16}
\int d^2\theta \b{D}^2 
\Lambda_{ijk^*l^*} 
(D^\alpha \Phi^i)( D_\alpha \Phi^j) 
(\b{D}_{\d\alpha} \b\Phi^{k^*})
(\b{D}_{\d\alpha} \b\Phi^{l^*})
+\hc,
\end{split}
\end{equation}
where $\Lambda_{ijk^*l^*}$ is a tensor 
$\Lambda_{ijk^*l^*}
=\Lambda_{ijk^*l^*}(\Phi^i,\b{\Phi}^{i^*}, \der_m\Phi^i,\der_m \b{\Phi}^{i^*},
D^2 \Phi^i,\b{D}^2 \b{\Phi}^{i^*}) $. 
If we consider multicomponent 
4-form field strength superfields as chiral superfields, 
we need the corresponding boundary term.
We can obtain the boundary term by the same procedure as the following discussion.

\subsubsection{SUSY ghost/tachyon-free higher derivative 
Lagrangian: boundary part}

Now, we specify the boundary term 
which corresponds to the higher derivative Lagrangian
in \er{gf3}.
As mentioned in Sec.~\ref{bh},
the boundary term which corresponds to the higher derivative term 
is needed.
To specify the boundary term, we determine the chiral superfield
for the boundary term $T_{\rm HD}$.
This chiral superfield is a generalization of 
$T$ and $T_{KW}$ in the cases in which the Lagrangians are 
given by the canonical kinetic term in \er{180825.1329}
and by a K\"ahler potential as well as a superpotential in 
\er{180718.1611}, respectively.

We consider the variation of the Lagrangian in \er{gf3} 
by 
the 3-form prepotential $X$:
\begin{equation}
\begin{split}
& \delta {\cal L}_\text{HD,SUSY}
\\
&= 
-\df{1}{128}
\Bigg( \int d^2\theta \b{D}^2
\bigg(
\pd{\Lambda}{Y} \(-\df{1}{4}\b{D}^2 \delta X\) |D_\alpha Y|^4
+
\pd{\Lambda}{\b{Y}} \(-\df{1}{4} D^2 \delta X\) |D_\alpha Y|^4
\\
&
\hph{-\df{1}{128} \Bigg(\int d^2\theta \b{D}^2\Big(}
\quad
+
\pd{\Lambda}{\der_m Y}  \(-\df{1}{4} \der_m \b{D}^2 \delta X\) |D_\alpha Y|^4
+
\pd{\Lambda}{\der_m \b{Y}}  
\(-\df{1}{4} \der_m D^2 \delta X\) |D_\alpha Y|^4
\\
&
\hph{-\df{1}{128} \Bigg(\int d^2\theta \b{D}^2\Big(}
\quad
+
\pd{\Lambda}{D^2 Y}  
\(-\df{1}{4} D^2 \b{D}^2 \delta X\) |D_\alpha Y|^4
+
\pd{\Lambda}{\b{D}^2 \b{Y}}  
\(-\df{1}{4} \b{D}^2 D^2 \delta X\) |D_\alpha Y|^4
\\
&
\hph{-\df{1}{128} \Bigg(\int d^2\theta \b{D}^2\Big(}
\quad
+
2\Lambda  (D^\alpha Y) \(-\df{1}{4}D_\alpha \b{D}^2 \delta X\)
(\b{D}_{\d\alpha} \b{Y})(\b{D}^{\d\alpha} \b{Y}) 
\\
&
\hph{-\df{1}{128} \Bigg(\int d^2\theta \b{D}^2\Big(}
\quad
+
2 \Lambda  (D^\alpha Y) (D_\alpha Y)
(\b{D}_{\d\alpha} \b{Y})
\(-\df{1}{4}\b{D}^{\d\alpha}D^2 \delta X\)
\bigg)+\hc\Bigg) 
\\
&
\quad
+\delta {\cal L}_\text{HD,SUSY,bound.},
\end{split} 
\label{180816.2016}
\end{equation}
where 
$|D_\alpha Y|^4:= 
(D^\alpha Y) (D_\alpha Y) (\b{D}_{\d\alpha}\b{Y}) 
(\b{D}^{\d\alpha} \b{Y})$.
 
We will use partial integrations for each 
spinor or vector derivative terms to find the chiral superfield 
$T_{\rm HD}$.
Unlike the cases of $T$ and $T_{KW}$,
there are 
spinor or vector derivatives on the variation terms  
such as $\b{D}_{\d\alpha} \b{D}^2 \delta X$,
$\der_m \b{D}^2 \delta X$ and $D^2 \b{D}^2 \delta X$.
Naively, we would ignore the boundary term 
for each of the partial integrations.
However, we should be careful when the 3-form gauge field is included
because the total derivative terms may give $\delta C_{mnp}$ 
at the boundary.
Therefore, we explicitly execute the partial integration.

For example, we show the following relation
\begin{equation}
\begin{split}
&-\df{1}{128}\int d^2\theta \b{D}^2
\pd{\Lambda}{D^2 Y}  
\(-\df{1}{4} D^2 \b{D}^2 \delta X\) |D_\alpha Y|^4
\\
&
=
-\df{1}{128}\int d^2\theta \b{D}^2
\(
D^2 
\pd{\Lambda}{D^2 Y}  |D_\alpha Y|^4
\)
\(-\df{1}{4}  \b{D}^2 \delta X\)  
\end{split}
\end{equation}
which may be the most complicated term in \er{180816.2016}
to derive the relation.
To show this relation, we change the integration 
$\int d^2\theta \b{D}^2  $ to $\int d^2\b\theta D^2$.
The change gives us the following total derivative term:
\begin{equation}
\begin{split}
&-\df{1}{128}\int d^2\theta \b{D}^2
\pd{\Lambda}{D^2 Y}  
\(-\df{1}{4} D^2 \b{D}^2 \delta X\) |D_\alpha Y|^4
\\
&
=
-\df{1}{128}
(-4i)
\der^{\d\beta\beta}\([D_\beta,\b{D}_{\d\beta}]
\pd{\Lambda}{D^2 Y}  
\(-\df{1}{4} D^2 \b{D}^2 \delta X\) |D_\alpha Y|^4\) |
\\
&\quad
-\df{1}{128}\int d^2\b\theta D^2
\pd{\Lambda}{D^2 Y}   |D_\alpha Y|^4
\(-\df{1}{4} D^2 \b{D}^2 \delta X\).
\end{split}
\label{180816.2046}
\end{equation}
Note that we
can safely execute the partial integration 
for the term $\int d^2\b\theta D^2...$ in the right hand side
by using the identity $D^2 D_\alpha =0$.
We now consider the total derivative term in \er{180816.2046}.
In the total derivative, there are terms proportional to
$D^2 \b{D}^2 \delta X||_\text{bound.}$, 
$\b{D}_{\d\beta} D^2 \b{D}^2 \delta X||_\text{bound.}$,
and
 $D_\beta \b{D}_{\d\beta} D^2 \b{D}^2 \delta X||_\text{bound.}$.
By the boundary conditions in 
\ers{180815.1853} and \eqref{180910.0412}, 
we conclude that the boundary term is equal to zero.
Therefore, \er{180816.2046} can be rewritten as
\begin{equation}
\begin{split}
&-\df{1}{128}\int d^2\theta \b{D}^2
\pd{\Lambda}{D^2 Y}  
\(-\df{1}{4} D^2 \b{D}^2 \delta X\) |D_\alpha Y|^4
\\
&
=
-\df{1}{128}
\int d^2\b\theta D^2
\(
D^2\pd{\Lambda}{D^2 Y}   |D_\alpha Y|^4
\)
\(-\df{1}{4} \b{D}^2 \delta X\).
\end{split}
\label{180816.2126}
\end{equation}
We again change the integration $\int d^2\b\theta D^2$
to $\int d^2\theta \b{D}^2$ as 
\begin{equation}
\begin{split}
&-\df{1}{128}\int d^2\theta \b{D}^2
\pd{\Lambda}{D^2 Y}  
\(-\df{1}{4} D^2 \b{D}^2 \delta X\) |D_\alpha Y|^4
\\
&
=
-\df{1}{128}\int d^2\theta \b{D}^2
\(
D^2\pd{\Lambda}{D^2 Y}   |D_\alpha Y|^4
\)
\(-\df{1}{4} \b{D}^2 \delta X\)
\\
&
\quad
-\df{1}{128}
(+4i) \der^{\d\beta \beta} 
\([D_\beta, \b{D}_{\d\beta}]
\(
D^2\pd{\Lambda}{D^2 Y}   |D_\alpha Y|^4
\)
\(-\df{1}{4} \b{D}^2 \delta X\)\)|.
\end{split}
\label{180816.2127}
\end{equation}
The total derivative term does not have the term 
including $\delta C_{mnp}$ at the boundary, and 
this total derivative term is zero.
By using the identity $\b{D}_{\d\beta} \b{D}^2 =0$, 
we finally obtain 
\begin{equation}
\begin{split}
&-\df{1}{128}
\int d^2\theta \b{D}^2
\pd{\Lambda}{D^2 Y}  
\(-\df{1}{4} D^2 \b{D}^2 \delta X\) |D_\alpha Y|^4
\\
&
=
-\df{1}{128}\int d^2\theta 
\(
\b{D}^2 D^2\pd{\Lambda}{D^2 Y}   |D_\alpha Y|^4
\)
\(-\df{1}{4} \b{D}^2 \delta X\).
\end{split}
\label{180816.2132}
\end{equation}
As a consequence, boundary terms  
given by the partial integrations are equal to zero.
Thus, the variation of the higher derivative Lagrangian 
can be written by using the following chiral superfield $T_\text{HD}$ 
\begin{equation}
\begin{split}
 T_\text{HD}
= 
-\df{1}{4}\b{D}^2\cdot \df{1}{16}
\Bigg(
&
\pd{\Lambda}{Y} |D_\alpha Y|^4
-\der_m \(\pd{\Lambda}{\der_m Y}|D_\alpha Y|^4 \)
\\
&
+D^2 \(\pd{\Lambda}{D^2 Y} |D_\alpha Y|^4 \)
-2 D^\alpha 
(\Lambda (D_\alpha Y)(\b{D}_{\d\alpha} \b{Y}) (\b{D}^{\d\alpha}\b{Y}))
\Bigg)
\end{split}
\end{equation}
as
\begin{equation}
 \delta {\cal L}_\text{HD,SUSY}
= 
\(
\int d^2\theta T_\text{HD} \(-\df{1}{4} \b{D}^2 \delta X\)
+\hc
\)
+
\delta {\cal L}_\text{HD,SUSY,bound.} 
\end{equation}
The variation has the same structure as the 
previous quadratic derivative case in \er{180718.2003}.
The superspace integration leads to the boundary term with 
$\delta C_{mnp}$:
\begin{equation}
\begin{split}
 \delta {\cal L}_\text{HD,SUSY}
&
= 
i T_{I,\text{HD}} 
\(\df{1}{16} (D^2 \b{D}^2 - \b{D}^2 D^2) \delta X\)|
+
\delta {\cal L}_\text{HD,SUSY,bound.}
\cdots
\\
&
=
i T_{I,\text{HD}} (-4i)
\(\df{1}{16} \der^{\d\alpha \alpha}[D_\alpha,\b{D}_{\d\alpha}] \delta X\)|
+
\delta {\cal L}_\text{HD,SUSY,bound.} 
+
\cdots
\\
&
=
+ \df{\sr{2}}{3!}
\der_m  \(T_{I,\text{HD}} |
\epsilon^{mnpq}
\delta C_{npq}
\)
+
\delta {\cal L}_\text{HD,SUSY,bound.}
+
\cdots. 
\end{split}
\label{180817.1600}
\end{equation}
Here, the ellipsis $\cdots$ denotes the terms 
that do not contribute to the boundary terms,
 and 
$T_{I,\text{HD}} = \tfrac{1}{2i} (T_\text{HD} - \b{T}_\text{HD})$. 
 To see the variation more concretely, 
we express $T_\text{HD}$ in terms of the component fields.
For simplicity, we focus on the bosonic term of $T_\text{HD}|$.
In the WZ gauge in \er{180910.0422}, 
the bosonic term of $T_\text{HD}$ can be 
calculated as 
\begin{equation}
\begin{split}
 T_\text{HD}|
&
= 
2\Lambda |{\cal F}|^2 \b{\cal F}
-2\Lambda \b{\cal F} \der^m y \der_m \b{y}
\\
&
\quad
-4 \pd{\Lambda}{D^2 Y} 
\(
|{\cal F}|^4 -2 \der^m y \der_m \b{y} |{\cal F}|^2
+(\der^m y \der_m y )(\der^n \b{y} \der_n \b{y} )
\)
\\
&
\quad
+\text{(fermions)}.
\end{split}
\end{equation}
Therefore, the imaginary part $T_{I, \rm HD}$ is 
\begin{equation}
\begin{split}
 T_{I, \rm HD}|
&
= 
\df{1}{\sr{2}}\Lambda (H^2 + F^2) F
-\sr{2}\Lambda(\der^m y \der_m \b{y})  F
\\
&
\quad
-4  \(\text{Im} \pd{\Lambda}{D^2 Y}\) 
\(
\df{1}{4}(H^2 + F^2)^2 - \der^m y \der_m \b{y} (H^2 + F^2)
+(\der^m y \der_m y )(\der^n \b{y} \der_n \b{y} )
\)
\\
&
\quad
+\text{(fermions)}.
\end{split} 
\end{equation}
We now consider the boundary term which cancels the 
variation at the boundary in \er{180817.1600}.
Such boundary term is given by the I-type integral as follows:
\begin{equation}
 {\cal L}_\text{HD,SUSY,bound.}
 = 
\df{i}{4}\(
\int d^2\theta \b{D}^2
-\int d^2\b\theta D^2 
\)
T_{I,\text{HD}} X.
\end{equation}
In the WZ gauge, the boundary term is expressed as 
\begin{equation}
 {\cal L}_\text{HD,SUSY,bound.}
=
 - \df{\sr{2}}{3!}
\der_m  \(T_{I,\text{HD}} |
\epsilon^{mnpq}
\delta C_{npq}
\).
\end{equation}
Therefore, this boundary term precisely cancels the variation 
in \er{180817.1600}.

By adding a K\"ahler potential and a superpotential, 
we obtain the most general Lagrangian of 
the 3-form gauge field with an arbitrary order of the field strength:
\begin{equation}
\begin{split}
& {\cal L}_{KW, \rm HD,SUSY}
\\
&
= 
-\df{1}{8}\int  d^2\theta \b{D}^2 K(Y,\b{Y})
+
\int d^2\theta W(Y)
\\
&\quad
-\df{1}{8\cdot 16} \int d^2 \theta 
\b{D}^2 \Lambda(Y,\b{Y}, \der_m Y, \der_m \b{Y},
D^2 Y, \b{D}^2 \b{Y}) (D^\alpha Y)(D_\alpha Y)
(\b{D}_{\d\alpha} \b{Y}) (\b{D}^{\d\alpha} \b{Y})
\\
&
\quad
+\df{i}{4}\int d^2\theta \b{D}^2 (T_{I, KW}+ T_{I,\rm HD})X
+\hc
\end{split}
\label{180901.1702}
\end{equation}

\subsubsection{EOM for 3-form gauge field}
Before closing this section, 
we write down the general EOM for the 3-form gauge field, 
which should be useful for applications.
We consider the most general Lagrangian 
in \er{180901.1702}.
The EOM for the 3-form prepotential can be derived 
by the variation of the Lagrangian in the bulk: 
\begin{equation}
\begin{split}
 0=
& 
-\df{1}{4}\b{D}^2 \pd{K}{Y} +\pd{W}{Y} 
-\df{1}{4\cdot 16} \b{D}^2 \pd{\Lambda}{Y} |D_\alpha Y|^4
+\df{1}{4\cdot 16} 
\b{D}^2 \der_m \pd{\Lambda}{\der_m Y} |D_\alpha Y|^4
\\
&
-\df{1}{4\cdot 16}
\b{D}^2 D^2 \pd{\Lambda}{D^2 Y} |D_\alpha Y|^4
+\df{1}{2\cdot 16} 
\b{D}^2 D^\alpha \Lambda 
(D_\alpha Y) (\b{D}_{\d\alpha} \b{Y}) (\b{D}^{\d\alpha} \b{Y})
+\hc
\end{split}
\label{180822.2006}
\end{equation}
As we have seen in Sec.~\ref{sc},
the $\theta= \b\theta=0$ component of this EOM corresponds to the EOM for the 
 the auxiliary field $H$.
In order to obtain the EOM for the 3-form gauge field, 
we consider a derivative $[D_\beta, \b{D}_{\d\beta}]$ 
on the both hand sides of \er{180822.2006}:
\begin{equation}
\begin{split}
 0=
-\df{1}{4} (+2i)\der_{\beta\d\beta}
\Bigg(
&\b{D}^2 \pd{K}{Y} -4 \pd{W}{Y} 
+\df{1}{16}\b{D}^2 \pd{\Lambda}{Y} |D_\alpha Y|^4
-\df{1}{16}\b{D}^2 \der_m \pd{\Lambda}{\der_m Y} |D_\alpha Y|^4
\\
&
\quad
+
\df{1}{16}\b{D}^2 D^2 \pd{\Lambda}{D^2 Y} |D_\alpha Y|^4
-
\df{1}{8}\b{D}^2 D^\alpha \Lambda 
(D_\alpha Y) (\b{D}_{\d\alpha} \b{Y}) (\b{D}^{\d\alpha} \b{Y})
\Bigg)
+\hc
\end{split}
\label{180822.2155}
\end{equation}
The EOM can be solved as follows:
\begin{equation}
\begin{split}
k = 
\df{i}{4}
\Bigg(
&\b{D}^2 \pd{K}{Y} -4 \pd{W}{Y} 
+\df{1}{16} \b{D}^2 \pd{\Lambda}{Y} |D_\alpha Y|^4
-\df{1}{16} \b{D}^2 \der_m \pd{\Lambda}{\der_m Y} |D_\alpha Y|^4
\\
&
\quad
+
\df{1}{16} \b{D}^2 D^2 \pd{\Lambda}{D^2 Y} |D_\alpha Y|^4
-\df{1}{8}
\b{D}^2 D^\alpha \Lambda 
(D_\alpha Y) (\b{D}_{\d\alpha} \b{Y}) (\b{D}^{\d\alpha} \b{Y})
\Bigg)
+\hc,
\end{split}
\label{180825.1428} 
\end{equation}
where $k$ is a constant which will be determined by 
the boundary conditions. 
By using \ers{180822.2006} and \eqref{180825.1428},
we obtain the following solution:
\begin{equation}
\begin{split}
 ik =  
& 
-\df{1}{4}\b{D}^2 \pd{K}{Y} +\pd{W}{Y} 
-\df{1}{4 \cdot 16} \b{D}^2 \pd{\Lambda}{Y} |D_\alpha Y|^4
+\df{1}{4\cdot 16} \b{D}^2 \der_m \pd{\Lambda}{\der_m Y} |D_\alpha Y|^4
\\
&
-\df{1}{4\cdot 16}
\b{D}^2 D^2 \pd{\Lambda}{D^2 Y} |D_\alpha Y|^4
+\df{1}{2\cdot 16} 
\b{D}^2 D^\alpha \Lambda 
(D_\alpha Y) (\b{D}_{\d\alpha} \b{Y}) (\b{D}^{\d\alpha} \b{Y}).
\end{split}
\label{180825.1434}
\end{equation}
The bosonic sector of the $\theta = \b\theta =0$ 
component of the above solution is 
expressed as
\begin{equation}
\begin{split}
 ik =  
& 
\pd{^2 K}{Y\der\b{Y}}| \b{\cal F}
 +\pd{W}{Y} |
\\
&
-4
\pd{\Lambda}{D^2 Y}|
(|{\cal F}|^4 -2 |{\cal F}|^2 (\der^m y \der_m \b{y})
+
(\der^m y \der_m y)(\der^n \b{y} \der_n \b{y} ))
\\
&
+2 \Lambda (|{\cal F}|^2\b{\cal F} - \b{\cal F} \der^m y\der_m \b{y}).
\end{split}
\label{180825.1435}
\end{equation}

To solve this equation, we need to give a concrete model of 
$K$, $W$, and $\Lambda$.
We will discuss such models and their solutions elsewhere.

\section{Summary and discussion}
\label{sum}
We have considered higher derivative extensions of 
3-form gauge theories in the both bosonic and SUSY cases.
For the bosonic case, we have shown that higher derivative terms given by derivatives on the field strength causes a tachyon 
as long as the canonical kinetic term exists.
We have also argued that the tachyon can also be a ghost, depending on models and parameters.
We have shown that 
there is neither a tachyon nor a ghost
if the higher derivative terms are given by functions of 
the field strength but not of the derivative of the field strength.
This is because the EOM is not changed from the case of 
the canonical kinetic term.
We have confirmed that 
previously known higher derivative Lagrangians~\cite{Dvali:2005an,Dvali:2005zk} fall into this class,
Then we have specified the boundary term 
which corresponds to the 
ghost/tachyon-free higher derivative 
Lagrangian of an arbitrary order of the field strength.
For the SUSY case, 
we have shown that 
a naive higher derivative extension of 
the SUSY 3-form gauge theory
may cause a tachyon as well as a ghost, 
as long as the canonical kinetic term exists.
Then we have presented the most general 
ghost/tachyon-free Lagrangian of 
an arbitrary order of the field strength, the corresponding 
boundary term, 
and EOM for the 3-form gauge field.

There can be several extensions and applications of our work.
One may apply the higher derivative theory in Sec.~\ref{bh} 
to the cosmological constant problem. 
Since 
the Lagrangian gives us a more general 
constant term given by $G(-c) + c G'(-c)$ 
than that of the canonical kinetic term $-\tfrac{1}{2}c^2$,
it may give a correction 
to the application to the cosmological constant problem.

We can extend the higher derivative Lagrangians to
include a topological coupling between 
a 3-form gauge field and a pseudo-scalar field, 
which will give us 
 the potential of the pseudo-scalar field. 
The general discussion between higher derivative terms and 
potentials is known for bosonic case in Ref.~\cite{Dvali:2005an},
and so 
the SUSY extension is possible by using our new ghost/tachyon-free Lagrangian.
Since coupling gives us a potential for the pseudo-scalar field,
it should be useful for inflationary models.

To apply the higher derivative Lagrangian to cosmology, 
we should embed the Lagrangian into SUGRA.
It may be straightforward since 3-form gauge theories in SUGRA are known in Refs.~\cite{Binetruy:1996xw, Farakos:2017jme}.

In this paper, we have shown no ghost in the bosonic sector 
of the SUSY case.
We expect that there is no 
fermionic ghost in \er{180901.1702} by SUSY transformations.
However, a concrete explanation of whether  
fermionic ghosts are absent 
or not is an open question.
We may discuss it along the line of Refs.~\cite{Kimura:2017gcy,Kimura:2018sfs}.

It would also be interesting to consider solutions of 
the EOM in \er{180825.1434}.
In the case where $\Lambda$ is a constant, it is known that there is 
a non-trivial solution of the auxiliary field 
(so-called a non-canonical branch), where the canonical kinetic term for the boson in the chiral superfield vanishes at on-shell.
Our higher derivative term would deform this solution. 

Higher derivative chiral superfield Lagrangians 
give several BPS equations for BPS solitons 
\cite{Nitta:2014pwa,Nitta:2015uba}.
It is interesting to discuss whether our SUSY higher derivative 
Lagrangian of the 3-form gauge field admit any BPS equations and their soliton solutions.

\subsection*{Acknowledgements}
The authors thank Toshiaki Fujimori and Keisuke Ohashi for 
discussions.
R.~Y.~also thanks Shuntaro Aoki, Tetsutaro Higaki,
Nobuyoshi Ohta and Yusuke Yamada for discussions. 
This work is supported by the Ministry of Education, 
Culture, Sports, Science (MEXT)-Supported
Program for the Strategic Research Foundation  at
Private Universities ``Topological Science'' (Grant No.~S1511006). 
M.~N.~is also supported in part by JSPS Grant-in-Aid
for Scientific Research (KAKENHI Grant No.~16H03984 and No.~18H01217),
and also by MEXT KAKENHI Grant-in-Aid for Scientific 
Research on Innovative Areas ``Topological Materials Science'' 
No.~15H05855.
\appendix
\section{Notation}
\label{notation}

In this appendix, we summarize our notation.
We use Wess--Bagger's notation in Ref.~\cite{Wess:1992cp}.
The Minkowski metric $\eta_{mn}$ 
and the totally anti-symmetric tensor $\epsilon_{mnpq}$
are given by
\begin{equation}
\eta_{mn} = (-1,1,1,1),
\quad 
 \epsilon_{0123} = -\epsilon^{0123} = 1,
\end{equation}
respectively.
Here, the Roman letters beginning with $m,n,p,...$ 
denote indices of vectors and tensors.

In the following, we summarize our notation of spinors.
The Greek letters beginning with $\alpha,\beta,...$
denote undotted spinor indices, 
while dotted Greek letters 
beginning with $\d\alpha,\d\beta,...$
denote dotted spinor indices.
The undotted and dotted spinors are related to each other 
by the Hermitian conjugate denoted by $\dg$.
The Hermitian conjugate of a spinor $\chi_\alpha$, 
 $\b\chi_{\d\alpha}$ are defined by
\begin{equation}
 (\chi_\alpha)^\dg = \b\chi_{\d\alpha}, 
\quad
 (\b\chi_{\d\alpha})^\dg = \chi_{\alpha}.
\end{equation}
The Hermitian conjugate of a product of spinors $\chi_\alpha$ 
and $\psi_\alpha$ is defined by
\begin{equation}
(\chi_\alpha\psi_\beta)^\dg
= \b\psi_{\d\beta} \b\chi_{\d\alpha}.
\end{equation} 
Spinor indices are raised and lowered 
by the following totally antisymmetric tensors
$\epsilon_{\alpha\beta}$, $\epsilon^{\alpha\beta}$,
 $\epsilon_{\d\alpha\d\beta}$, and $\epsilon^{\d\alpha\d\beta}$
with the following normalizations
\begin{equation}
 \epsilon^{12} = - \epsilon^{21} 
= -\epsilon_{12} = \epsilon_{21} = 1,
\quad
 \epsilon^{\d{1}\d{2}} = - \epsilon^{\d{2}\d{1}} 
=  -\epsilon_{\d{1}\d{2}} =  \epsilon_{\d{2}\d{1}} = 1
\end{equation}
as
\begin{equation}
 \chi^\alpha = \epsilon^{\alpha\beta} \chi_\beta,
\quad
 \chi_\alpha = \epsilon_{\alpha\beta} \chi^\beta,
\quad
 \b\chi^{\d\alpha} = \epsilon^{\d\alpha\d\beta} \b\chi_{\d\beta},
\quad
 \b\chi_{\d\alpha} = \epsilon_{\d\alpha\d\beta} \b\chi^{\d\beta}.
\end{equation}
The anti-symmetric tensors satisfy
\begin{equation}
\epsilon^{\alpha\beta}\epsilon_{\beta\gamma} 
= \delta^\alpha_\gamma,
\quad
\epsilon^{\d\alpha\d\beta}\epsilon_{\d\beta\d\gamma} 
= \delta^{\d\alpha}_{\d\gamma}, 
\end{equation}
where $\delta^\alpha_\beta$ and $\delta^{\d\alpha}_{\d\beta}$ 
are the Kronecker's delta for undotted and dotted 
spinors, respectively.
The contraction of spinors are 
\begin{equation}
 \psi^\alpha \chi_\alpha 
 = \epsilon^{\alpha\beta} \psi_\beta \chi_\alpha,
\quad
 \b\psi_{\d\alpha} \b\chi^{\d\alpha} 
 = \epsilon^{\d\alpha\d\beta} \b\psi_{\d\beta}\b\chi^{\d\alpha}.
\end{equation}

A vector is represented by a tensor product of spinors.
The relations between vectors and spinors are 
given by the 4D Pauli matrices $(\sigma_m)_{\alpha\d\alpha}$
defined by 
\begin{equation}
 (\sigma_m)_{\alpha\d\beta}
=(\sigma_0, \sigma_1,\sigma_2,\sigma_3)_{\alpha\d\beta}
 = \(
\mtx{1 & 0 \\ 0 & 1},
\mtx{0 & 1 \\ 1 & 0},
\mtx{0 & -i \\ i & 0},
\mtx{1 & 0 \\ 0 & -1}
\).
\end{equation}
For example, the spacetime derivative $\der_m $ are 
represented by using spinors as
\begin{equation}
\der_{\alpha\d\alpha} = (\sigma^m)_{\alpha\d\alpha}\der_m.
\end{equation}
The relation of Hermitian conjugates of the Pauli matrices 
$(\b\sigma_m)^{\d\alpha \beta} $ to $(\sigma_m)_{\alpha\d\beta}$ is
\begin{equation}
 (\b\sigma_m)^{\d\alpha \beta}
=
 (\sigma_m)^{\beta\d\alpha}
= 
\epsilon^{\d\alpha\d\gamma}\epsilon^{\beta\delta}
 (\sigma_m)_{\delta \d\gamma}.
\end{equation}
The Pauli matrices satisfy the following relation:
\begin{equation}
(\sigma_m)_{\alpha\d\beta} (\b\sigma_n)^{\d\beta\gamma}
+ 
(\sigma_n)_{\alpha\d\beta} (\b\sigma_m)^{\d\beta\gamma}
=
-2\eta_{mn}\delta_\alpha^\gamma,
\quad
(\b\sigma_m)^{\d\alpha\beta} (\sigma_n)_{\beta\d\gamma}
+ 
(\b\sigma_n)^{\d\alpha\beta} (\sigma_m)_{\beta\d\gamma}
=
-2\eta_{mn}\delta^{\d\alpha}_{\d\gamma}.
\end{equation}
The matrices $\sigma_{mn} $ and $\b\sigma_{mn}$ are 
defined by
\begin{equation}
 \begin{split}
 (\sigma_{mn})_\alpha{}^\beta 
 &= 
\df{1}{4} ((\sigma_m)_{\alpha\d\gamma}(\b\sigma_n)^{\d\gamma \beta}
-(\sigma_n)_{\alpha\d\gamma}(\b\sigma_m)^{\d\gamma \beta}),
\\ 
 (\b\sigma_{mn})^{\d\alpha}{}_{\d\beta} 
& = 
\df{1}{4} ((\b\sigma_m)^{\d\alpha\gamma}(\sigma_n)_{\gamma \d\beta}
-(\sigma_n)^{\d\alpha\gamma}(\b\sigma_m)_{\gamma \d\beta}). 
 \end{split}
\end{equation}

To formulate SUSY theories, we have used SUSY covariant spinor 
derivatives.
The definition of the spinor derivatives are
\begin{equation}
 D_\alpha = \pd{}{\theta^\alpha}
+ i\b\theta^{\d\beta} (\sigma_m)_{\alpha\d\beta} \pd{}{x^m},\quad
\b{D}^{\d\alpha}
 = -\pd{}{\b\theta_{\d\alpha}}
-i\theta_\beta (\b\sigma_m)^{\d\alpha\beta} \pd{}{x^m}.
\end{equation}
\section{Duality transformation of bosonic 3-form gauge theory}
\label{bd}

In this Appendix, 
we review the duality transformation of the 3-form.
The 3-form is classically dual to a constant.
This can be shown as follows.
We consider the following first order Lagrangian 
instead of the Lagrangian in \er{3}:
\begin{equation}
\begin{split}
  {\cal L}'_\text{kin.} 
&= 
-\df{1}{2\cdot 4!} F'^{mnpq}F'_{mnpq}
+\df{1}{3!} \der^m (C^{npq}F'_{mnpq})
+ \df{1}{4!} q \epsilon^{mnpq} ( F'_{mnpq}  - 3 \der_m C_{npq}).
\end{split}
\label{4}
\end{equation}
Here, $F'_{mnpq}$ is a 4-form field which is independent 
of the 3-form gauge field $C_{mnp}$, but we assume the boundary 
condition for the 4-form:
\begin{equation}
 F' = -c,
\end{equation}
where $F' = \tfrac{1}{4!}\epsilon^{mnpq}F'_{mnpq}$.
In \er{4}, $q$ is a 
pseudo-scalar field which is regarded as a Lagrange's multiplier.
The vanishing of the variation by the 3-form gauge field
at the boundary requires the boundary condition for $q$:
\begin{equation}
 q|_\text{bound.} = c.
\end{equation}
The EOM of $q$ gives 
$\epsilon^{mnpq} F'_{mnpq} = 4 \epsilon^{mnpq} \der_m C_{npq}$,
and we obtain the original Lagrangian in \er{3}.
On the other hand, the EOM for the 4-form field $F'_{mnpq}$ gives 
\begin{equation}
 F'_{mnpq} =  q \epsilon_{mnpq}, \qtq{or equilvalently}
q = -F'.
\end{equation}
Substituting the solution to the Lagrangian in \er{4}, 
we obtain 
\begin{equation}
\begin{split}
 {\cal L}'_\text{kin.} 
&= 
-\df{1}{2} q^2
+\df{1}{3!} \der^m (C^{npq}\epsilon_{mnpq} q)
- \df{1}{3!} q \epsilon^{mnpq} \der_m C_{npq}
\\
&
=-\df{1}{2} q^2
+\df{1}{3!} C^{npq}\epsilon_{mnpq} \der^m q.
\end{split}
\end{equation}
The EOM of $C_{mnp}$ gives 
\begin{equation}
 \der_m q=0.
\end{equation}
Therefore, $q$ is local constant, which is equal to $c$ by
the boundary condition.
Therefore, the Lagrangian is equal to a constant.
\begin{equation}
 {\cal L}'_\text{kin.} 
= 
-\df{1}{2} c^2.
\end{equation}

The reverse of the duality transformation is
possible, and the boundary term in \er{3} 
is naturally understood by this transformation.
We start with the following Lagrangian:
\begin{equation}
{\cal L}_\text{const.}= -\df{1}{2} c^2,
\end{equation}
where $c$ is a real constant.
The first-order Lagrangian is 
\begin{equation}
 {\cal L}'_\text{const.} = -\df{1}{2} q^2 + 
\df{1}{3!}\epsilon^{mnpq}C_{npq} \der_m q,
\label{180827.1804}
\end{equation}
where we assume the boundary conditions for $q$ as
$q|_\text{bound.} = c$.
The Lagrangian can be constructed as follows.
The constant term can be considered as a closed 0-form.
A closed 0-form $f$ is defined by the condition $\der_m f=0$.
Thus, we introduce the condition as a solution of the Lagrange's 
multiplier field.
In 4D, the Lagrange's multiplier 
can be a 3-form, and this condition can be imposed by the term
 $\epsilon^{mnpq} C_{npq} \der_m q$.
In this case, there is a gauge symmetry of $C_{mnp}$ given in \er{5}.

The EOM for $q$ is 
\begin{equation}
 q= -\df{1}{4!} \epsilon^{mnpq} \der_m C_{npq} = 
-\df{1}{4!} \epsilon^{mnpq} F_{mnpq} = -F.
\end{equation}
Substituting this solution into \er{180827.1804}, 
we obtain
\begin{equation}
\begin{split}
 {\cal L}'_\text{const.} 
&
= -\df{1}{2} F^2 -
\df{1}{3!}\epsilon^{mnpq}C_{npq} \der_m F
= +\df{1}{2} F^2 -
\df{1}{3!} \der_m (\epsilon^{mnpq}C_{npq} F)
\\
&
= -\df{1}{2\cdot 4!} F^{mnpq}F_{mnpq} +
\df{1}{3!} \der_m (C_{npq} F^{mnpq}).
\end{split}
\label{180827.1811} 
\end{equation}
Thus, we have the Lagrangian in \er{3}.
\section{Duality transformation of SUSY 3-form gauge theory}
\label{sd}
In this Appendix, we review a duality transformation of 
SUSY 3-form gauge theories~\cite{Farakos:2017jme}.
A 3-from field in SUSY theories can be dualized 
into a chiral superfield $\Phi$
which has a linear superpotential $ir\Phi$
with a real constant $r$.
The boundary term for the 3-form gauge theories
can be straightforwardly obtained 
by this duality procedure.
Here, we only consider the case of
 a canonical kinetic term, 
although this discussion can be generalized 
to the case in which the Lagrangian is given by 
a K\"ahler potential and a superpotential.

We consider a dual transformation from a single chiral superfield $\Phi$
 with a linear superpotential to a 3-form field system.
We start with the following Lagrangian:
\begin{equation}
 {\cal L}_\text{kin.,chiral}=-\df{1}{8}\int d^2\theta \b{D}^2 
\Phi\b\Phi 
+\int  d^2\theta ir\Phi +\hc,
\label{170714.2201}
\end{equation}
where $r$ is a real constant.

Now we dualize the Lagrangian by considering the following 
Lagrangian with a chiral superfield $Q$
and a real superfield $X$:
\begin{equation}
 {\cal L}'_\text{kin.,chiral}
=
-\df{1}{8}\int d^2\theta \b{D}^2 
\Phi\b\Phi 
+
\int d^2\theta  Q \Phi
+\df{1}{8}\int  d^2\theta \b{D}^2 
X (Q +\b{Q})
+\hc,
\label{170714.2211}
\end{equation}
where we assume the boundary condition $Q||_\text{bound.} = ir$.
Note that $X$ has a gauge symmetry $X \to X + L$, where 
$L$ is a linear superfield $D^2 L =\b{D}^2 L = 0$.
The gauge transformation implies that $X$ is a 3-form prepotential.
The EOM of $X$ gives us $Q+\b{Q}=0$,
i.e.~$Q$ is a pure imaginary constant, and $Q$ is equal to $ir$
by the boundary condition.
Substituting this solution into \er{170714.2211}, 
we obtain the original Lagrangian in \er{170714.2201}.

Instead, the EOM for the chiral superfield $Q$ relates
the original chiral superfield $\Phi$ with the real superfield $X$:
\begin{equation}
\Phi = -\df{1}{4} \b{D}^2 X =: Y,
\end{equation}
where $Y$ can be identified with a 4-form field strength 
superfield since $Y$ is related to the real superfield $X$.
The EOM for the chiral superfield $\Phi$ leads to 
the relation between $Y$ and $Q$:
\begin{equation}
 -\df{1}{4}\b{D}^2 \b\Phi = -\df{1}{4} \b{D}^2 \b{Y}= -Q.
\end{equation}
The solution gives us the boundary condition for $D^2 Y$:
\begin{equation}
 -\df{1}{4}D^2 Y||_\text{bound.}
=-\b{Q} ||_\text{bound.} = ir.
\end{equation}
Eliminating the chiral superfield $\Phi$,
 we obtain the Lagrangian 
\begin{equation}
 {\cal L}'_\text{kin.,chiral}
=
-\df{1}{8}\int d^2\theta \b{D}^2 
Y\b{Y}
+\df{i}{4}\int  d^2\theta \b{D}^2 
X T_I
+\hc,
\label{180827.2120}
\end{equation}
Here $T_I$ is a imaginary part of the chiral superfield 
$T := -\tfrac{1}{4} \b{D}^2 \b{Y}$.
The second term in the right hand side in \er{180827.2120}
is the boundary term which is equal to the one in 
\er{180718.1541}.
Note that to derive the Lagrangian in \er{180827.2120}, 
we have used the following calculation:
\begin{equation}
\begin{split}
& 
\int d^2\theta  Q \Phi
+\df{1}{8}\int  d^2\theta \b{D}^2 
X (Q +\b{Q})
=
-\df{1}{4}\int d^2\theta \b{D}^2  X Q  
+\df{1}{8}\int  d^2\theta \b{D}^2 
X (Q +\b{Q})
\\
 &
=
\df{i}{4}\int  d^2\theta \b{D}^2 
X \df{1}{2i}\( -\df{1}{4}\b{D}^2 \b{Y}+ \df{1}{4}D^2 Y\).
\end{split} 
\end{equation}
The Lagrangian in \er{180827.2120} is the same as the one in 
\er{180825.1329}, and thus the Lagrangian with 
a linear superpotential in \er{170714.2201} 
can be dualized into a Lagrangian for a 3-form gauge field
including the boundary term.

\section{Ghost/tachyon-free Lagrangian and auxiliary field method}
\label{sgf}

Here, we consider an auxiliary field method for 
the ghost/tachyon-free Lagrangian proposed in Sec.~\ref{sh}.
In Sec.~\ref{sh}, we have shown that the Lagrangian in 
\er{180901.1702} is ghost/tachyon-free by using component expression.
We can also show it by using the auxiliary field method
in Sec.~\ref{sg}.

The Lagrangian with 
auxiliary superfields is given as follows:
\begin{equation}
\begin{split}
& {\cal L}'_\text{HD,SUSY}
\\
& = 
\Bigg(
\df{1}{8\cdot 16}
\int d^2 \theta \b{D}^2\Lambda(Y,\b{Y}, \der_m Y, \der_m \b{Y},
\b\Psi, \Psi) (D^\alpha Y)(D_\alpha Y)
(\b{D}_{\d\alpha} \b{Y}) (\b{D}^{\d\alpha} \b{Y})
\\
&
\quad\hph{\quad\Bigg(}
+
\int d^2\theta \Sigma \(\Psi +\df{1}{4} \b{D}^2 \b{Y}\)
+\hc\Bigg) 
+{\cal L}'_{\rm HD,SUSY,bound.}.
\end{split}
\label{180814.1825}
\end{equation}
Here, ${\cal L}'_{\rm HD,SUSY,bound.}$ is the boundary term
for this Lagrangian, which is not relevant to 
the following discussion.
In this Lagrangian, the superfield $\Sigma$ is not an 
independent dynamical superfield 
as far as fermions are set to be zero in the vacuum.
This can be seen by the EOM for the chiral superfield $\Psi$.
The EOM for $\Psi$ is 
\begin{equation}
 0
= 
-\tfrac{1}{4} \b{D}^2 
\(
\pd{\Lambda}{\Psi} 
(D^\alpha Y)(D_\alpha Y)
(\b{D}_{\d\alpha} \b{Y}) (\b{D}^{\d\alpha} \b{Y}) 
\)
+
\Sigma.
\end{equation}
The EOM leads to $\Sigma| =0$ 
around the vacuum where $D_\alpha Y|=0$.
Thus, $\Sigma|$ has
no dynamical degrees of freedom.
Therefore, $\Sigma|$ does not give rise to a ghost.
As long as SUSY is preserved, the fermionic 
component $D_\alpha \Sigma |$ is also not dynamical, 
and so there is no fermionic ghost as well.

%\bibliography{./../../../../yokokura}
\providecommand{\href}[2]{#2}

\end{document}